\documentclass[prx,twocolumn]{revtex4-2}
\usepackage{amsmath,amssymb} 
\usepackage{graphicx} 
\usepackage{hyperref} 
\usepackage{bm} 
\usepackage{float} 
\usepackage[final]{changes} 
\usepackage{booktabs} 
\usepackage{cleveref} 
\usepackage{xcolor} 
\usepackage{mathtools} 
\usepackage{lmodern} 
\usepackage[T1]{fontenc} 
\usepackage{setspace} 
\usepackage{array} 
\usepackage{tikz} 
\usepackage{soul}
\usepackage{comment}
\usepackage{lineno}

\renewcommand{\figurename}{Fig}

\usepackage{tikz}
\usetikzlibrary{tikzmark}

\Crefname{equation}{Eq.}{Eqs.}

\graphicspath{{FIG/}}

\def\d{\text{d}}
\begin{document}
\title{Reconstructing Noisy Gene Regulation Dynamics Using
  Extrinsic-Noise-Driven Neural Stochastic Differential Equations}


\author{Jiancheng Zhang$^{1\#}$, Xiangting Li$^{2\#}$, Xiaolu Guo$^{3\# *}$, Zhaoyi You$^4$, Lucas B\"ottcher$^{5,6}$, Alex Mogilner$^7$, Alexander Hoffmann$^3$, Tom Chou$^{2,8*}$, Mingtao Xia$^{7, 9*}$}
\affiliation{
  $^1$Department of Electrical and Computer Engineering, University of California, Riverside, CA, USA. 
  $^2$Department of Computational Medicine, University of California, Los Angeles, CA, USA. \\
  $^3$Department of Microbiology, Immunology, and Molecular Genetics (MIMG), \& Institute for Quantitative and Computational
Biosciences, University of California Los Angeles, Los Angeles, CA, USA. \\
  $^4$Ray and Stephanie Lane Computational Biology Department, School of Computer Science, Carnegie Mellon University, Pittsburgh, PA, USA. \\
  $^5$Department of Computational Science and Philosophy, Frankfurt School of Finance and Management, Frankfurt am Main, Germany. \\
  $^6$Laboratory for Systems Medicine, Department of Medicine, University of Florida, FL 32610, USA. \\
  $^7$Courant Institute of Mathematical Sciences, New York University, NY 10012, USA. \\
  $^8$Department of Mathematics, University of California, Los Angeles, CA 90095, USA.\\
  $^9$Department of Mathematics, University of Houston, Houston,
            TX 77004, USA.
  ($^\#$indicates equal contribution)
}
\email{xiaoluguo@g.ucla.edu, tomchou@ucla.edu, xiamingtao@nyu.edu}


\begin{abstract}
  Proper regulation of cell signaling and gene expression is crucial
  for maintaining cellular function, development, and adaptation to
  environmental changes.  \added{Reaction dynamics in cell populations
    is often noisy} because of (i) inherent stochasticity of
  intracellular biochemical reactions (``intrinsic noise'') and (ii)
  heterogeneity of cellular states across different cells that are
  influenced by external factors (``extrinsic noise''). In this work,
  we introduce an extrinsic-noise-driven neural stochastic
  differential equation (END-nSDE) framework that utilizes the
  Wasserstein distance to accurately reconstruct SDEs from trajectory
  data \added{from a heterogeneous population of cells (extrinsic
    noise).}  We demonstrate the effectiveness of our approach using
  both simulated and experimental data from three different systems in
  cell biology: (i) circadian rhythms, (ii) RPA-DNA binding
  dynamics, and (iii) NF$\kappa$B signaling process. Our END-nSDE
  reconstruction method can model how cellular heterogeneity
  (extrinsic noise) modulates reaction dynamics in the presence of
  intrinsic noise. It also outperforms existing time-series analysis
  methods such as recurrent neural networks (RNNs) and long short-term
  memory networks (LSTMs). By inferring cellular heterogeneities from
  data, our END-nSDE reconstruction method can reproduce noisy
  dynamics observed in experiments. In summary, the reconstruction
  method we propose offers a useful surrogate modeling approach for
  complex biophysical processes, where high-fidelity mechanistic models
  may be impractical.
\end{abstract}
  
\maketitle

\section{Introduction}
\added{Reactions that control signaling and gene regulation} are
important for maintaining cellular function, development, and
adaptation to environmental changes, which impact all aspects of
biological systems, from embryonic development to an organism's
ability to sense and respond to environmental signals. Variations in
gene regulation, arising from noisy biochemical
processes~\cite{Swain2002,Michael2002}, can result in phenotypic
heterogeneity even in a population of genetically identical
cells~\cite{sanchez2013regulation}.

Noise within cell populations \added{can be categorized as} (i)
``intrinsic noise,'' which arises from the inherent stochasticity of
biochemical reactions and quantifies, \textit{e.g.}, biological
variability across cells in the same state
\cite{Michael2002,Foreman2020,Mitchell2018}, and (ii) ``extrinsic
noise,'' which encompasses heterogeneities in environmental factors or
differences \added{in cell state across a population.}  A substantial
body of literature has focused on quantifying intrinsic and extrinsic
noise from experimental and statistical
perspectives~\cite{thattai2001intrinsic,Swain2002,Michael2002,tsimring2014noise,Fu2016,Llamosi2016February,Dharmarajan2019January,finkenstadt2013quantifying,dixit2013quantifying,fang2024advanced}. Experimental
studies have specifically identified relevant sources of noise in
various organisms, including \textit{E. coli} (Escherichia coli),
yeast, and mammalian
systems~\cite{Michael2002,Raj2006,Raser2004,Sigal2006,Singh2012}.

Extrinsic noise is associated with uncertainties in biological
parameters that vary across different cells. The distribution over
physical and chemical parameters determine the observed variations in
cell states, concentrations, locations of regulatory proteins and
polymerases \cite{Michael2002,Swain2002,Paulsson2005}, and
transcription and translation rates \cite{Singh2014}. For example,
extrinsic noise is the main contributor to the variability of
concentrations of oscillating p53 protein levels \added{across cell
  populations \cite{wang2019roles}}.
%
%
On the other hand, intrinsic noise, \textit{i.e.}, inherent
stochasticity of cells in the same state, can limit the accuracy of
expression and signal
transmission~\cite{Michael2002,Mitchell2018}. Based on the law of mass
action~\cite{Voit2015,Ferner2015}, ordinary differential equations
(ODEs) apply only in some deterministic or averaged limit and do not
take into account intrinsic noise. Therefore, stochastic models are
necessary to accurately represent biological processes, such as
thermodynamic fluctuations inherent to molecular interactions within
regulatory networks~\cite{Mitchell2018,Paulsson2005,Swain2002} or
random event times in birth-death processes.
%
%

\added{Existing stochastic modeling methods that account for intrinsic
  noise include Markov jump processes \cite{Paul2013,Kurtz1976} and
  SDEs \cite{TIAN2007,Chen2005,Kinetic_gene2024}.}  Additionally, a
hierarchical Markov model was designed in \cite{zechner2014} for
parameter inference in dual-reporter experiments to separate the
contributions of extrinsic noise, intrinsic noise, and measurement
error when both extrinsic and intrinsic noise are present. The
described methods have been effective in the reconstruction of
low-dimensional noisy biological systems. However, these methods
usually require specific forms of a stochastic model with unknown
parameters to be inferred. It is unclear whether these methods and
their generalizations can be applied to more complex (\textit{e.g.},
higher-dimensional) systems for which a mechanistic description of the
underlying biophysical dynamics is not available or impractical.

\begin{figure*}
  \raggedright
    \centering 
        \includegraphics[width=5in]{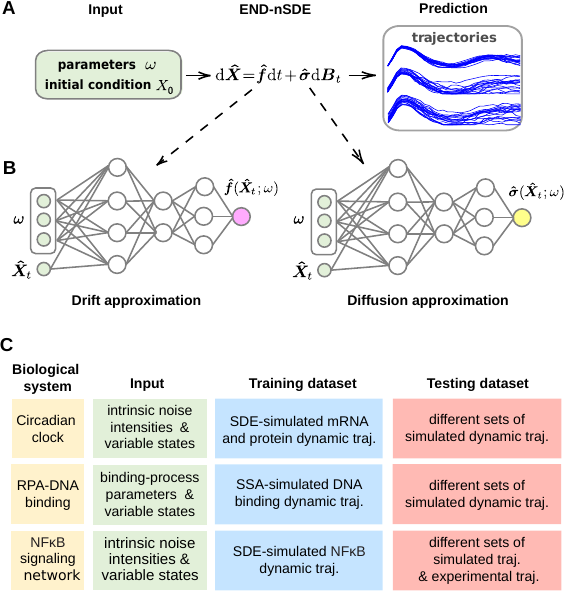}
    \caption{\textbf{Workflow of our proposed END-nSDE prediction on
        parameters altering stochastic dynamics.} A. The predicted
      trajectories are generated through the reconstructed SDE
      $\mathrm{d}{\hat{\bm{X}}} =
      {\hat{\bm{f}}}({\hat{\bm{X}}};\omega)\mathrm{d}t +
      \hat{\bm{\sigma}}(\hat{\bm{X}};\omega)\mathrm{d}\bm{B}_t$. B. The
      drift and diffusion functions, $\hat{\bm{f}}$ and
      $\hat{\bm{\sigma}}$, are approximated using parameterized neural
      networks. The parameterized neural-network-based drift function
      $\hat{\bm{f}}({\hat{\bm{X}}};\omega)$ and diffusion function
      $\hat{\bm{\sigma}}({\hat{\bm{X}}};\omega)$ take the system state
      ${\hat{\bm{X}}}$ and biological parameters $\omega$ as
      inputs. C. Diagram of three examples illustrating the nSDE
      input, along with training and testing datasets.}
    	\label{fig:frame}
\end{figure*}

In this work, we introduce an extrinsic-noise-driven neural stochastic
differential equation (END-nSDE) reconstruction method. Our method
builds upon a recently developed Wasserstein distance ($W_{2}$
distance) nSDE reconstruction method~\cite{xia2024squared} to identify
macromolecular reaction kinetics and cell signaling dynamics from
noisy observational data under the presence of both extrinsic and
intrinsic noise. \added{A key question we address in this paper is how
  extrinsic noise that characterizes cellular heterogeneity influences
  the overall stochastic dynamics of the population.} Our approach
differs from the one proposed in Ref.~\cite{xia2024squared} in that
our method takes into account both cell heterogeneity and intrinsic
fluctuations.  In Fig.~\ref{fig:frame}, we provide an overview of the
systems that we study in this work.

Our approach employs neural networks as SDE approximators in
conjunction with the \texttt{torchsde} package~\cite{li2020,
  kidger2021} for reconstructing noisy dynamics from data.  Previous
work showed that for SDE reconstruction tasks, the $W_{2}$ distance
nSDE reconstruction method outperforms other benchmark methods such as
generative adversarial
networks~\cite{xia2024squared,kidger2021neural}.  Additionally, the
$W_{2}$ distance nSDE reconstruction method can directly extract the
underlying SDE from temporal trajectories without requiring specific
mathematical forms of the terms in the underlying SDE model.
\deleted{Our proposed END-nSDE incorporates extrinsic noise which is
  critical for understanding heterogeneity in the intrinsically
  stochastic dynamics.}  We apply our END-nSDE methodology to three
biological processes that illustrate (i) circadian clock rhythm, (ii)
RPA-DNA binding dynamics, and (iii) NF$\kappa$B signaling to show that
END-nSDE can predict how extrinsic noise modulates stochastic dynamics
with intrinsic noise. Additionally, our method demonstrates superior
performance compared to several time-series modeling methods including
recurrent neural networks (RNNs), long short-term memory networks
(LSTMs), and Gaussian processes. In summary, the reconstruction method
we propose provides a useful surrogate modeling approach for complex
biomedical processes, especially in scenarios where high-fidelity
mechanistic models are impractical.

\section{Methods and Models}
\label{method}
In this work, we extend the temporally decoupled squared
$W_2$-distance SDE reconstruction method proposed in
Refs.~\cite{xia2024squared,xia2024efficient} to reconstruct
\added{noisy dynamics across a heterogeneous cell population}
(``extrinsic noise''). Our goal is to not only reconstruct SDEs for
approximating noisy cellular signaling dynamics from time-series
experimental data, but to also quantify how heterogeneous biological
parameters, such as enzyme- or kinase-mediated biochemical reaction
rates, affect such noisy cellular signaling dynamics.


\subsection{SDE reconstruction with heterogeneities in biological parameters}
The $W_{2}$-distance-based neural SDE reconstruction method proposed in
Ref.~\cite{xia2024squared} aims to approximate the SDE
\begin{equation}~\label{eq:grtruthsde}
  \mathrm{d} \bm{X}(t)=  \bm{f} (\bm{X}(t) ,t)\mathrm{d}t+
  \bm{\sigma} (\bm{X}(t) ,t)\mathrm{d}\mathbf{B}(t),\,\, \bm{X}(t)\in\mathbb{R}^d,
\end{equation}
using an approximated SDE
\begin{equation}~\label{eq:approxtruthsde}
  \mathrm{d} \hat{\bm{X}}(t)=  \hat{\bm{f}}(\hat{\bm{X}}(t) ,t)\mathrm{d}t
  +\hat{\bm{\sigma}} (\hat{\bm{X}}(t) ,t)\mathrm{d}\bm{B}(t),\,\,
  \hat{\bm{X}}(t)\in\mathbb{R}^d,
\end{equation}
where $\bm{\hat{f}}$ and $\bm{\hat{\sigma}}$ are two parameterized
neural networks that approximate the drift and diffusion functions
$\bm{f}$ and $\bm{\sigma}$ in Eq.~\eqref{eq:grtruthsde},
respectively. These two neural networks are trained by minimizing a
temporally decoupled squared $W_2$-distance loss function
\begin{equation}\label{loss}
  \tilde{W}_{2}^{2}(\mu ,\hat{\mu})=
  \!\int_0^T\!\!\!
  \inf_{\pi (\mu(t) ,\hat{\mu}(t))}\!\!\mathbb{E} _{\pi (\mu(t) ,\hat{\mu}(t))}
  \Big[ \big\| \bm{X}(t)-\hat{\bm{X}}(t) \big\|^{2} \Big] \text{d} t,
\end{equation}
where $\bm{X}(t)$ and $\hat{\bm{X}}(t)$ are the observed trajectories
at time $t$ and trajectories generated by the approximate SDE model
Eq.~\eqref{eq:approxtruthsde} at time $t$, respectively.
$\mathbb{E}_{\pi (\mu(t) ,\hat{\mu}(t))}\Big[ \big\| \bm{X}(t)
  -\hat{\bm{X}}(t) \big\|^{2} \Big]$ represents the expectation when
$(\bm{X},\hat{\bm{X}})\sim \pi (\mu(t) ,\hat{\mu}(t))$.  Our
$W_{2}$-distance-based SDE reconstruction method can result in very small
errors $\bm{f}-\bm{\hat{f}}$ and $\bm{\sigma}-\bm{\hat{\sigma}}$ in
the reconstructed diffusion and jump functions.  The $\tilde{W}_{2}^2$
term in Eq.~\eqref{loss} is denoted as the squared temporally
decoupled squared $W_2$ distance loss function.  For simplicity, in
this paper, we shall also denote the Eq.~\eqref{loss} as the squared
$W_2$ loss. $\mu$ and $\hat{\mu}$ are the probability distributions
associated with the stochastic processes $\{\bm{X}(t)\}, 0\leq t\leq
T$ and $\{\hat{\bm{X}}(t)\}, 0\leq t\leq T$, respectively, while
$\mu(t)$ and $\hat{\mu}(t)$ are the probability distributions of
$\bm{X}(t)$ and $\hat{\bm{X}}(t)$ at a specific time $t$. The coupling
distributions $\pi$ of two distributions $\mu(t), \hat{\mu}(t)$ on the
probability space $\mathbb{R}^d$ are defined by
\begin{equation}
\begin{aligned}
        \pi (A, \mathbb{R}^d) = \mu(A), \,\,
        \pi (\mathbb{R}^d, B) = \hat{\mu}(B),\,\, \forall A, B\in\mathcal{B}(\mathbb{R}^d),
\end{aligned}
\label{coupling_def}
\end{equation}
where $\mathcal{B}(\mathbb{R}^d)$ is the Borel $\sigma$-algebra on
$\mathbb{R}^d$.  The infimum in Eq.~\eqref{loss} is taken over all
possible coupling distributions $\pi(\mu(t), \hat{\mu}(t))$ and
$\|\cdot\|$ denotes the $\ell^2$ norm of a vector. That is,
\begin{equation}
    \big \| \bm{X}(t) \big \|^2\coloneqq \sum_{i=1}^{d}  \big| X_{i}(t) \big|^{2}.
\end{equation}

Across different cells, extrinsic noise or cellular heterogeneities
such as differences in kinase or enzyme abundances resulting from
cellular variabilities, can lead to variable, cell-specific, gene
regulatory dynamics. Such heterogeneous and stochastic gene
expression (both intrinsic and extrinsic noise) can be modeled using
SDEs with distributions of parameter values reflecting cellular
heterogeneity. To address heterogeneities in gene dynamics across
different cells, we propose an END-nSDE method that is able to
reconstruct a family of SDEs for the same gene expression process
under different parameters.  Specifically, for a given set of
(biological) parameters $\omega$, we are interested in reconstructing
\begin{equation}
  ~\label{eq:grtruthsde_para}
  \mathrm{d} \bm{X}(t;\omega)=
  \bm{f} (\bm{X}(t;\omega);\omega)\mathrm{d}t
  +\bm{\sigma} (\bm{X}(t;\omega);\omega)
  \mathrm{d}\bm{B}(t),
\end{equation}
using the approximate SDE
\begin{equation}~\label{eq:approxtruthsde_para}
  \mathrm{d} \hat{\bm{X}}(t;\omega)=
  \hat{\bm{f}}(\hat{\bm{X}}(t;\omega);\omega)
  \mathrm{d}t+\hat{\bm{\sigma}} (\hat{\bm{X}}(t;\omega);\omega)
  \mathrm{d}\hat{\bm{B}}(t),
\end{equation}
in the sense that the errors $\bm{f}(\bm{X}(t;\omega);\omega) -
\hat{\bm{f}} (\bm{X}(t;\omega);\omega)$ and $\bm{\sigma}(\bm{X}(t;\omega);\omega) - \hat{\bm{\sigma}} (\bm{X}(t;\omega);\omega)$ for all
different values of $\omega$ can be minimized. In
Eq.~\eqref{eq:approxtruthsde_para}, $\hat{\bm{f}}$ and
$\hat{\bm{\sigma}}$ are represented by two parameterized neural
networks that take both the state variable $\hat{\bm{X}}$ and the
parameters $\omega$ as inputs. To train these two neural networks, we
propose an extrinsic-noise-driven temporally decoupled squared $W_2$
distance loss function
\begin{equation}~\label{eq:W_2}
  L(\Lambda) = \sum_{\omega\in \Lambda}
  \tilde{W}_2^2(\mu(\omega), \hat{\mu}(\omega) ),
\end{equation}
where $\mu(\omega)$ and $\hat{\mu}(\omega)$ are the distributions of
the trajectories $\bm{X}(t;\omega),\, 0\leq t\leq T$ and
$\hat{\bm{X}}(t;\omega),\, 0\leq t\leq T$, and $\tilde{W}$ is the
temporally decoupled squared $W_2$ loss function in Eq.~\eqref{loss}.
$\Lambda$ denotes the set of parameters $\omega$. Note that
Eq.~\eqref{eq:W_2} is different from the local squared $W_{2}$ loss
in Refs. \cite{xia2024local, xia2025newlocaltimedecoupledsquared} since we do not require a continuous
dependence of $\{\bm{X}(t;\omega)\}_{t\in[0, T]}$ on the parameter $\omega$
nor do we require that $\omega$ is a continuous variable. The
extrinsic-noise-driven temporally decoupled squared $W_2$ loss
function Eq.~\eqref{eq:W_2} takes into account both parameter
heterogeneity and intrinsic fluctuations as a result of the Wiener
process $\bm{B}(t)$ and $\hat{\bm{B}}(t)$ in
Eqs.~\eqref{eq:grtruthsde} and \eqref{eq:approxtruthsde}.

We summarize the END-nSDE method in Figs.~\ref{fig:frame}A, B. With
observed noisy single-cell dynamic trajectories as the training data,
we train two parameterized neural networks by minimizing
Eq.~\eqref{eq:W_2} to approximate the drift and diffusion terms in
the SDE. The reconstructed nSDE is a surrogate model of single-cell
dynamics (see Figs.~\ref{fig:frame} A, B).  The hyperparameters and
settings for training the neural SDE model are summarized in
Table~\ref{tab:all-sde} of Appendix~\ref{sec:sdeap}.

\subsection{Biological models}
\label{models}
We consider three biological examples where stochastic dynamics play a
critical role and use our END-nSDE method to reconstruct noisy
single-cell gene expression dynamics under both intrinsic and
extrinsic noise (also summarized in Fig.~\ref{fig:frame}C). In these
applications, we investigate the extent to which the END-nSDE can
efficiently capture and infer changes in the dynamics driven by
extrinsic noise.

\subsubsection{Noisy oscillatory circadian clock model}
Circadian clocks, often with a typical period of approximately 24
hours, are ubiquitous in intrinsically noisy biological rhythms
generated at the single-cell molecular level
\cite{Gonze2011September}.

We consider a minimal SDE model of the periodic gene \deleted{regulatory}
dynamics responsible for \textit{per} gene expression which is
critical in the circadian cycle. Since \textit{per} gene
expression is subject to intrinsic noise \cite{Westermark2009}, we
describe it using a linear damped-oscillator SDE
\begin{equation}
	\begin{aligned}
	  & \d x=-\alpha x{\d t} - \beta y {\d t}+ \xi_{x1}\d{B_{1,t}}
          + \xi_{x2}\d{B_{2,t}}\\
	  & \d y=\beta x{\d t} - \alpha y {\d t}+\xi_{y1}\d{B_{1,t}}
          + \xi_{y2}\d{B_{2,t}},
	\end{aligned}
 \label{example1_model}
\end{equation}
where $x$ and $y$ are the dimensionless concentrations of the
\textit{per} mRNA transcript and the corresponding \textit{per}
protein, respectively. $\d{B_{1,t}}$, $\d{B_{2,t}}$ are two
independent Wiener processes and the parameters $\alpha>0$ and
$\beta>0$ denote the damping rate and angular frequency,
respectively. A stability analysis at the steady state $(x, y)=(0, 0)$
in the noise-free case ($\xi_x=\xi_y=0$ in Eq.~\eqref{example1_model})
reveals that the real parts of the eigenvalues of the Jacobian matrix
$\begin{psmallmatrix} -\alpha & -\beta \\ \beta &
  -\alpha \end{psmallmatrix}$ at $(x, y)=(0, 0)$ are all negative,
indicating that the origin is a stable steady state when the system is
noise-free. Noise prevents the state $(x(t), y(t))$ from staying at
$(0, 0)$; thus, fluctuations in the single-cell circadian rhythm is
noise-induced \cite{Westermark2009}.

To showcase the effectiveness of our proposed END-nSDE method, we take
different forms of the diffusion functions $\xi_x$ and $\xi_y$ in
Eq.~\eqref{example1_model}, accompanied by different values of noise
strength and the correlation between the diffusion functions in the
dynamics of $x, y$.

\subsubsection{RPA-DNA binding model} \label{ex_dna}
Regulation of gene expression relies on complex interactions between
proteins and DNA, often described by the kinetics of binding
and dissociation.  Replication protein A (RPA) plays a pivotal role in
various DNA metabolic pathways, including DNA replication and repair,
through its dynamic binding with single-stranded DNA (ssDNA)
\cite{Dueva2020September,Caldwell2020August,Nguyen2017March,Wold1997June}. By
modulating the accessibility of ssDNA, RPA regulates multiple
biological mechanisms and functions, acting as a critical regulator
within the cell~\cite{qi2023ssdna}. Understanding the dynamics of
RPA-ssDNA binding is therefore a research area of considerable
biological interest and significance.

\begin{figure}[htbp]
    \centering 
    \includegraphics[width=\linewidth]{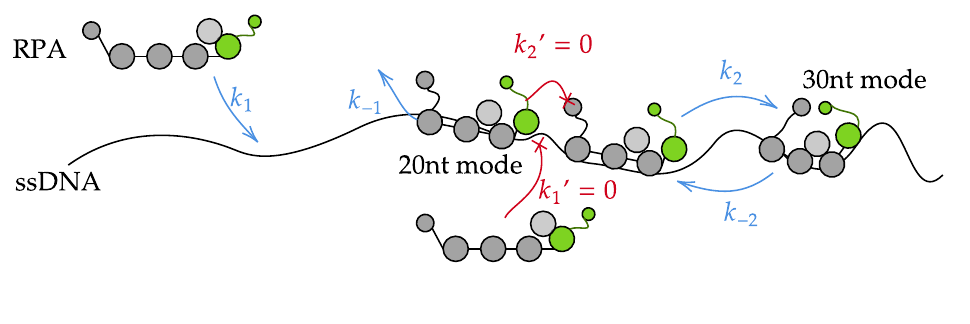}
    \caption{\textbf{A continuous-time discrete Markov chain model for
        multiple RPA molecules binding to long ssDNA.}  The possible
      steps in the \added{biomolecular kinetics} of RPA on ssDNA
      create a complex scenario involving multiple RPA molecules
      binding.  The RPA in the free solution can bind to ssDNA with
      rate $k_{1}$ provided there are at least 20 nucleotides (nt) of
      consecutive unoccupied sites.  This bound ``20-nt mode'' RPA
      unbinds with rate $k_{-1}$. When space permits, the 20nt-mode
      RPA can extend and bind an additional 10nt of DNA at a rate of
      $k_2$, converting it to a 30nt mode bound protein. The 30nt-mode
      RPA transforms back to 20nt-mode spontaneously with the rate
      $k_{-2}$. However, when the gap is not large enough to
      accommodate the RPA, the binding or conversion is prohibited
      ($k_{1}'=0$ and $k_{2}'=0$).}
    \label{fig:4ab}
\end{figure}

Multiple binding modes and volume exclusion effects complicate the
modeling of RPA-ssDNA dynamics.  The RPA first binds to ssDNA in 20
nucleotide (nt) mode, which occupies 20nt of the ssDNA.  When the
subsequent 10nt of ssDNA is free, 20nt-mode RPA can transform to
30nt-mode, further stabilizing its binding to ssDNA, as illustrated in
Fig.~\ref{fig:4ab}.  Occupied ssDNA is not available for other
proteins to bind.  Consequently, the gap size between adjacent
ssDNA-bound RPAs determines the ssDNA accessibility to other proteins.

Mean-field mass-action type chemical kinetic ODE models cannot
describe the process very well because they do not capture the
intrinsic stochasticity.  A stochastic model that tracks the fraction
of two different binding modes of RPA, 20nt-mode ($x_1$) and 30nt-mode
($x_2$), has been developed to capture the dynamics of this process.
A brute-force approach using forward stochastic simulation algorithms
(SSAs)~\cite{Gillespie1977dec} was then used to fit the model to
experimental data~\cite{qi2023ssdna}. However, a key challenge in this
approach is that the model is nondifferentiable with respect to the
kinetic parameters, making it difficult to estimate parameters.  Yet,
simple spatially homogeneous stochastic chemical reaction systems can
be well approximated by a corresponding SDE of the form given in
Eq.~\eqref{eq:grtruthsde} when the variables are properly scaled in
the large system size limit~\cite{Gillespie2000jul}. While
interparticle interactions shown in Fig.~\ref{fig:4ab} make it
difficult to find a closed-form SDE approximation, the results in
Ref.~\cite{Gillespie2000jul} motivate the possibility of an SDE
approximation for the RPA-ssDNA binding model in terms of the
variables $x_1$ and $x_2$.

Here, to address the non-differentiability issue associated with the
underlying Markov process, we use our END-nSDE model to construct a
differentiable surrogate for SSAs, allowing it to be readily trained
from data. Further details on the models and data used in this study
are provided in Appendix~\ref{sec:dna_binding}. Throughout our
analysis of RPA-DNA binding dynamics, we benchmark the SDE
reconstructed by our extended $W_{2}$-distance approach against those
found using other time series analysis and reconstruction methods such
as the Gaussian process, RNN, LSTM, and the neural ODE model. We
show that our surrogate SDE model is most suitable for approximating
the RPA-DNA binding process because it can capture the intrinsic
stochasticity in the dynamics.
\subsubsection{NF$\kappa$B signaling model}
\label{nfkbmodel}
Macrophages can sense environmental information and respond
accordingly with stimulus-response specificity encoded in signaling
pathways and decoded by downstream gene expression profiles
\cite{sheu2019stimulus}.  The temporal dynamics of NF$\kappa$B, a key
transcription factor in immune response and inflammation, encodes
stimulus information \cite{adelaja2021six}.  NF$\kappa$B targets and
regulates vast immune-related genes
\cite{hoffmann2002ikappab,cheng2021nf,sen2020gene}.  While NF$\kappa$B
signaling dynamics are stimulus-specific, they exhibit significant
heterogeneity across individual cells under identical conditions
\cite{adelaja2021six}. Understanding how specific cellular
heterogeneity (extrinsic noise) contributes to heterogeneity in
NF$\kappa$B signaling dynamics can provide insight into how noise
affects the fidelity of signal transduction in immune cells.

\begin{figure}[htbp]
    \centering 
    \includegraphics[width=1.8in]{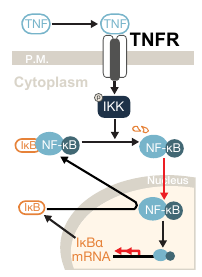}
    \caption{\textbf{Simplified schematic of the NF$\kappa$B Signaling
        Network.} TNF binds its receptor, activating IKK, which
      degrades I$\kappa$B$\alpha$ and releases NF$\kappa$B. The free
      NF$\kappa$B translocates to the nucleus and promotes
      I$\kappa$B$\alpha$ transcription. Newly synthesized
      I$\kappa$B$\alpha$ then binds NF$\kappa$B and exports it back to
      the cytoplasm. Red arrows indicate noise that we consider in the
      corresponding SDE system.}
    \label{fig:nfkb-scheme}
\end{figure}

A previous modeling approach employs a 52-dimensional ODE system to
quantify the NF$\kappa$B signaling network \cite{adelaja2021six} and
recapitulate the signaling dynamics of a representative cell. This ODE
model includes 52 molecular entities and 47 reactions across a
TNF-receptor module, an adaptor module, and a core module with and
NF$\kappa$B-IKK-I$\kappa$B$\alpha$ (I$\kappa$B$\alpha$ is an inhibitor
of NF$\kappa$B, while IKK is the I$\kappa$B kinase complex that
regulates the I$\kappa$B$\alpha$ degradation) feedback loop (see
Fig.~\ref{fig:nfkb-scheme}) \cite{adelaja2021}.  However, such an ODE
model is deterministic and assumes no intrinsic fluctuations in the
biomolecular processes. Yet, from experimental data, the NF$\kappa$B
signaling dynamics fluctuate strongly; such fluctuations cannot be
quantitatively described by any deterministic ODE model. Due to the
system's high dimensionality and nonlinearity, it is challenging to
quantify how intrinsic noise influences temporal coding in NF$\kappa$B
dynamics.

To incorporate the intrinsic noise within the NF$\kappa$B signaling
network, we introduce noise terms into the 52-dimensional ODE system
to build an SDE that can account for the observed temporally
fluctuating nuclear NF$\kappa$B trajectories.  While NF$\kappa$B
signaling pathways involve many variables, experimental constraints
limit the number of measurable components. Among these, nuclear
NF$\kappa$B activity is the most direct and critical experimental
readout. As a minimal stochastic model, we hypothesize that only the
biophysical and biochemical processes of NF$\kappa$B translocation
(which directly affects experimental measurements) and
I$\kappa$B$\alpha$ transcription (a key regulator of NF$\kappa$B
translocation) are subject to Brownian-type noise (red arrows in
Fig.~\ref{fig:nfkb-scheme}), as these processes play crucial roles in
the oscillatory dynamics of NF$\kappa$B \cite{adelaja2021}.

The intensity of Brownian-type noise in the NF$\kappa$B dynamics may
depend on factors such as cell volume (smaller volumes result in
higher noise intensity), or copy number (lower copy numbers lead to
greater noise intensity), and is therefore considered a form of
extrinsic noise. Noise intensity parameters thus capture an aspect of
cellular heterogeneity. There are other sources of cellular
heterogeneity, such as variations in kinase or enzyme abundances,
which are too complicated to model and are thus not included in the
current model. For simplicity, all kinetic parameters, except for the
noise intensity ($\sigma$), are assumed to be consistent with those of
a representative cell \cite{adelaja2021}. The 52-dimensional ODE model
for describing NF$\kappa$B dynamics is given in
Refs.~\cite{Guo2024Modeling,adelaja2021six}.  We extend this model by
adding noise to the dynamics of the sixth, ninth, and tenth ODEs of
the 52-dimensional ODE model. We retain 49 ODEs but convert the
equations for the sixth, ninth, and tenth components to SDEs:
\begin{widetext}
    \begin{equation}
\begin{aligned}
  \d u_6 & = \Big(k_{\text{basal}} + \frac{k_{\text{max}}
    u_{52}^{n_{{\text{NF}\kappa
          \text{B}}}}}{u_{52}^{n_{\text{NF}\kappa \text{B}}} +
    K_{\text{NF}\kappa\text{B}}^{n_{\text{NF}\kappa\text{B}}} }- k_{\text{deg}} u_6 \Big)\d t
  + \sigma_1 \d{B_{1,t}}\\[6pt]
  \d u_{9} & =  \big(k_{\text{imp}} u_{9} - k_{\text{a-I}\kappa\text{B-NF}\kappa\text{B}}
  u_{2}  u_{9} - k_{\text{deg-NF}\kappa\text{B}} u_{9}+ v^{-1} k_{\text{exp}}u_{10}
  + k_{\text{d-I}\kappa\text{B-NF}\kappa\text{B}} u_{4} + k_{\text{phos}}
  u_{7}\big)\d t - \sigma_2 \d{B_{2,t}}\\[8pt]
    \d u_{10} &=  \big(- k_{\text{exp}} u_{10} - k_{\text{a-I}\kappa\text{B-NF}\kappa\text{B}} u_3 u_{10} + v k_{\text{imp}} u_9+ k_{\text{d-I}\kappa\text{B-NF}\kappa\text{B}} u_5\big)\d t
+ \sigma_2 \d{B_{2,t}}.
\end{aligned}
\label{eq:nfkb_model-3}
\end{equation}
\end{widetext}
In Eqs. \ref{eq:nfkb_model-3}, $u_{2}$ is the concentration of
I$\kappa$B$\alpha$; $u_{3}$ is the concentration of
I$\kappa$B$\alpha$; $u_{4}$ is the concentration of the
I$\kappa$B$\alpha$-NF$\kappa$B complex; $u_{5}$ is the concentration
of the I$\kappa$B$\alpha$-NF$\kappa$B complex in the nucleus; $u_6$ is
the mRNA of I$\kappa$B$\alpha$; $u_7$ is the
IKK-I$\kappa$B$\alpha$-NF$\kappa$B complex; $u_9$ is NF$\kappa$B;
$u_{10}$ represents nuclear NF$\kappa$B activity; and $u_{52}$ is the
nuclear concentration of NF$\kappa$B with RNA polymerase II that is
ready to initiate mRNA transcription. A description of the parameters
and their typical values are given in the supplemental table
\ref{tab:nfkb-para}.  The quantities $\sigma_1\d{B}_{1,t}$ and
$\sigma_2\d{B}_{2,t}$ are noise terms associated with
I$\kappa$B$\alpha$ transcription and NF$\kappa$B translocation,
respectively. The remaining variables are latent variables and their
dynamics are regulated via the remaining 49-dimensional ODE in
Refs.~\cite{Guo2024Modeling,adelaja2021six}. The activation of
NF$\kappa$B is quantified by the nuclear NF$\kappa$B concentration
($u_5 + u_{10}$), which is also measured in experiments.

Within this example, we wish to determine if our proposed
parameter-associated nSDE can accurately reconstruct the dynamics
underlying experimentally observed NF$\kappa$B trajectory data.

\section{Results}
\label{results}
\subsection{Accurate reconstruction of circadian clock dynamics}
As an illustrative example, we use the $W_{2}$-distance nSDE
reconstruction method to first reconstruct the minimal model for
damped oscillatory circadian dynamics (see Eq.~\eqref{example1_model})
under different forms of the diffusion function. We set the two
parameters $\alpha=0.19$ and $\beta= 0.21$ in
Eq.~\eqref{example1_model} and impose three different forms for the
diffusion functions $\xi_{x1}, \xi_{x2}, \xi_{y1}, \xi_{y2}$: a
constant diffusion function \cite{bressloff2014stochastic}, a Langevin
\cite{spagnolo2009noise} diffusion function, and a linear diffusion
function \cite{pahle2012biochemical}. These functions, often used to
describe fluctuating biophysical processes, are
\begin{equation}~\label{eq:constxy}
 \text{const:} \begin{bmatrix}\xi_{x1} & \xi_{x2}
 \\\xi_{y1} & \xi_{y2}
\end{bmatrix}=  \sigma_{0}  
\begin{bmatrix}
1 & c \\
c & 1 
\end{bmatrix}\,,
\end{equation}
\begin{equation}~\label{eq:langxy}
  \text{Langevin:} \begin{bmatrix}\xi_{x1} & \xi_{x2}
 \\\xi_{y1} & \xi_{y2}
\end{bmatrix}= \sigma_{0}   
\begin{bmatrix}
\sqrt{\left |x \right |} & c \sqrt{\left |y \right |} \\[4pt]
c \sqrt{\left |x \right |} & \sqrt{\left |y \right |}
\end{bmatrix}\,,
\end{equation}
and
\begin{equation}~\label{eq:linearxy}
  \text{linear:} \begin{bmatrix}\xi_{x1} & \xi_{x2}
 \\\xi_{y1} & \xi_{y2}
\end{bmatrix}= \sigma_{0}   
\begin{bmatrix} 
x & c \left | y \right | \\
c \left | x \right | &  y
\end{bmatrix}\,.
\end{equation}
There are two additional parameters in Eqs.~\eqref{eq:constxy},
\eqref{eq:langxy}, and ~\eqref{eq:linearxy}: $\sigma_0$ that
determines the intensity of the Brownian-type fluctuations and $c$
that controls the correlation of fluctuations between the two
dimensions. For each type of diffusion function, we trained a
different nSDE model, each of which takes the state variables $(x, y)$
and the two parameters $(c, \sigma_0)$ as inputs and which outputs the
values of the reconstructed drift and diffusion functions.


We take 25 combinations of $(\sigma_0, c)\in\big\{(0.1+0.05i,
0.2+0.2j), i\in \{0,...,4\}, j\in \{0,...,4\}\big\}$; for each
combination of $(\xi_{x1}, \xi_{x2}, \xi_{y1}, \xi_{y2})$, we generate
50 trajectories from the ground truth SDE~\eqref{example1_model} as
the training data with $t\in[0, 1]$. The initial condition is set as
$(x(0), y(0))= (0, 1)$. To test the accuracy of the
reconstructed diffusion and drift functions, we measure the following
relative errors:
\begin{widetext}
    \begin{equation}~\label{eq:totaleror}
    \text{Error in } \bm{f} \coloneqq \frac{\sum_{i=1}^M \sum_{j=0}^{T}\lvert \bm{f} (\bm{X}_i(t_j;\omega) ;\omega)\mathrm-\hat{\bm{f}} (\hat{\bm{X}}_i(t_j;\omega) ;\omega)\mathrm  \rvert_1 }{\sum_{j=0}^{T}\lvert  \bm{f} (\bm{X}_i(t_j;\omega) ;\omega)  \rvert_1 },
\end{equation}
\begin{equation}~\label{eq:totaleror2}
    \text{Error in } \bm{\sigma} \coloneqq 
 \frac{ \sum_{i=1}^M\sum_{j=0}^{T}  \lvert |\bm{\sigma}(\bm{X}_i(t_j;\omega);\omega) \bm{\sigma}^T(\bm{X}_i(t_j;\omega), t_j;\omega)| - |\bm{\hat{\sigma}}(\bm{X}_i(t_j;\omega);\omega) \bm{\hat{\sigma}}^T(t_j;\omega)|  \rvert_{\text{m}}}{\sum_{i=0}^{M} \sum_{j=0}^{d} \lvert \bm{\sigma}(\bm{X}_i(t_j;\omega);\omega) \bm{\sigma}^T(\bm{X}_i(t_j;\omega);\omega) \rvert_{\text{m}}}.
\end{equation}

\end{widetext}


%
Here, $\bm{f}\coloneqq(-\alpha x - \beta y, \beta x-\alpha y)^T$ is
the vector of ground truth drift functions and $\hat{\bm{f}}$ is the
reconstructed drift function. $\bm{\sigma}$ is the matrix of ground
truth diffusion functions $[\xi_{x1}, \xi_{x2}; \xi_{y1}, \xi_{y2}]$
given in Eqs.~\eqref{eq:constxy}, \eqref{eq:langxy}, and
\eqref{eq:linearxy}. $M$ is the number of training samples,
$|\cdot|_1$ denotes the $\ell^1$ norm of a vector, and the matrix norm $|A|_{\text{m}}\coloneqq \sum_{i=1}^m\sum_{j=1}^n|A_{ij}|$
for a matrix $A\in\mathbb{R}^{m\times n}$. The errors are measured
separately for different parameters $\omega\coloneqq (\sigma_0, c)$.

\begin{figure*}
	\centering
		 	\includegraphics[width=5.2in]{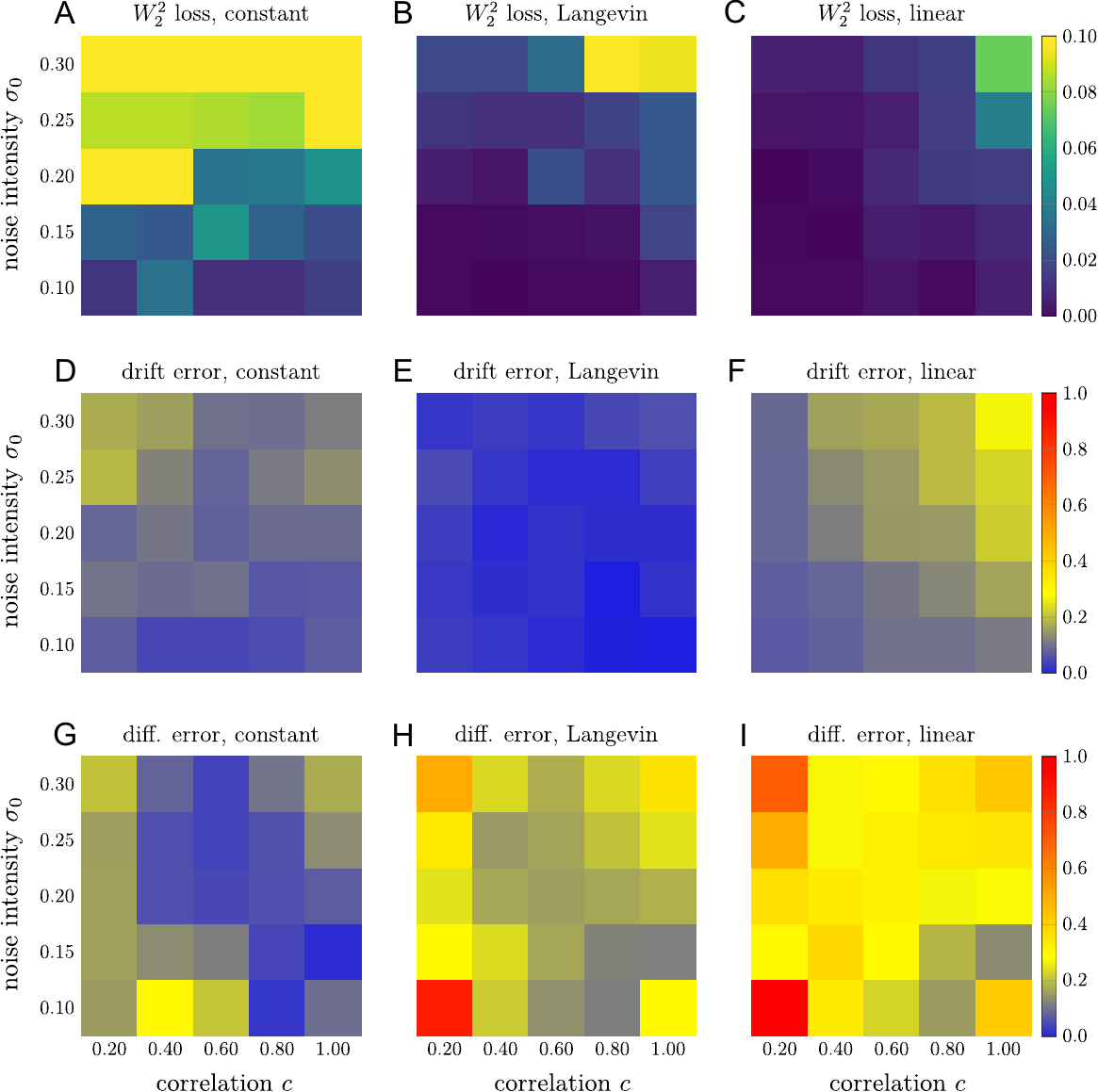}
	                \caption{\textbf{Reconstructing the circadian
                            model using END-nSDE.} Temporally
                          decoupled squared $W_2$ losses
                          Eq.~\eqref{loss} and errors in the
                          reconstructed drift and diffusion functions
                          for different types of the diffusion
                          function and different values of $(\sigma_0,
                          c)$. A-C. The temporally decoupled squared
                          $W_2$ loss between the ground truth
                          trajectories and the trajectories generated
                          by the reconstructed nSDEs for the
                          constant-type diffusion function
                          Eq.~\eqref{eq:constxy}, Langevin-type
                          diffusion function Eq.~\eqref{eq:langxy},
                          and the linear-type diffusion function
                          Eq.~\eqref{eq:linearxy}.  D-F. Errors in the
                          reconstructed drift function for the three
                          different types of ground truth diffusion
                          functions and the linear-type diffusion
                          function
                          Eq.~\eqref{eq:linearxy}. G-I. Errors in the
                          reconstructed diffusion function for the
                          three different types of ground truth
                          diffusion functions.}
	\label{fig:error}
\end{figure*}

The errors in the reconstructed drift function $\hat{\bm{f}}$ and
diffusion function $\hat{\bm{\sigma}}$ as well as the temporally
decoupled squared $W_2$ loss Eq.~\eqref{loss} associated with
different forms of the diffusion function and different values of
$(\sigma_0, c)$ are shown in Fig.~\ref{fig:error}. When the diffusion
function is a constant Eq.~\eqref{eq:constxy}, the mean reconstruction
error of the drift function is 0.15, the mean reconstruction error of
the diffusion function is 0.16, and the mean temporally decoupled
squared $W_2$ loss between the ground truth trajectories and the
predicted trajectories is 0.074 (averaged over all sets of parameters
$(\sigma_0, c)$). When a Langevin-type diffusion function
Eq.~\eqref{eq:langxy} is used as the ground truth, the mean errors for
the reconstructed drift and diffusion functions are 0.069 and 0.29,
respectively, and the mean temporally decoupled squared $W_{2}$ loss
between the ground truth and predicted trajectories is 0.020. For a
linear-type diffusion function as the ground truth, mean
reconstruction errors of the drift and diffusion functions are 0.19
and 0.41, respectively, and the mean temporally decoupled squared
$W_{2}$ distance is 0.013. For all three forms of diffusion, our
END-nSDE method can accurately reconstruct the drift function
$(-\alpha x-\beta y, \beta x - \alpha y)$ (see
Figs.~\ref{fig:error}D-F). When the diffusion function is a constant,
our END-nSDE model can also accurately reconstruct this constant (see
Fig.~\ref{fig:error}G). When the diffusion function takes a more
complicated form such as the Langevin-type diffusion function
Eq.~\eqref{eq:langxy} or the linear-type diffusion function
Eq.~\eqref{eq:linearxy}, the reconstructed nSDE model can still
approximate the diffusion function well for most combinations of $(c,
\sigma_0)$, especially when the correlation $c>0.2$ (see
Figs.~\ref{fig:error}H-I). Overall, our proposed END-nSDE model can
accurately reconstruct the minimal stochastic circadian dynamical
model Eq.~\eqref{example1_model} in the presence of extrinsic noise
(different values of $(\sigma_0, c)$); the accuracy of the
reconstructed drift and diffusion functions is maintained for most
combinations of $(\sigma_0, c)$.


\subsection{Accurate approximation of interacting DNA-protein systems
  with different kinetic parameters}
\label{MC}
To construct a differentiable surrogate for stochastic simulation
algorithms (SSAs), the neural SDE model should be able to take kinetic
parameters as additional inputs. Thus, the original $W_{2}$-distance SDE
reconstruction method in \cite{xia2024squared} can no longer be
applied because the trained neural SDE model cannot take into account
extrinsic noise, \textit{i.e.}, different values of kinetic
parameters.  To be specific, we vary one parameter (the conversion rate $k_2$
from 20nt-mode RPA to 30nt-mode RPA) in the stochastic model and then
apply our END-nSDE method which takes the state variables and the
kinetic parameter $k_2$ as the input. We set $k_2\in\{10^{-4+j/10},
j=0,...,25\}$ with other parameters taken from
experiments~\cite{qi2023ssdna} ($k_{1}=10^{-3}$ s$^{-1}$, $k_{-1} =
10^{-6}$ s$^{-1}$, $k_{-2} = 10^{-6}$ s$^{-1}$, see
Fig.~\ref{fig:4ab}). For each $k_2$, we generate 100 trajectories and
use 50 for the training set and the other 50 for the testing set.
Each trajectory encodes the dynamics of the fraction of 20-nt mode
DNA-bound RPA $x_1(t)$ and the fraction of 30-nt mode DNA-bound RPA
$x_2(t)$.

When approximating the dynamics underlying the RPA-DNA binding
process, we compare our SDE reconstruction method with other benchmark
time-series analysis or reconstruction approaches, including the RNN,
LSTM, Gaussian process, and the neural ODE model. These benchmarks are
described in detail in Appendix \ref{sec:lstm}.
%


\begin{table}[ht]
\renewcommand{\arraystretch}{1.2}
\centering
\caption{The extrinsic-noise-driven time-decoupled squared $W_2$
  distance Eq.~\eqref{eq:W_2} between the ground truth and predicted
  trajectories generated by different models on the testing set.}
\begin{tabular}{lc}
\toprule
\multicolumn{1}{c}{\textbf{Model}} &
\multicolumn{1}{c}{\textbf{Loss}}\\
\hline
\textbf{END-nSDE} & \textbf{0.0006}
	\\

LSTM & 0.062\\
RNN & 0.087\\
nODE	& 0.0012 \\
Gaussian Process & 0.0010
	\\
\bottomrule
\end{tabular}
\label{tab:dna1}
\end{table}

The extrinsic-noise-driven temporally decoupled squared $W_2$ distance
loss Eq.~\eqref{eq:W_2} between the distribution of the ground truth
trajectories and the distribution of the predicted trajectories
generated by our END-nSDE reconstructed SDE model is the smallest
among all methods (shown in Table~\ref{tab:dna1}). The underlying
reason is an SDE well approximates the genuine Markov counting process
process underlying the continuum-limit RPA-DNA binding process
\cite{Gillespie2000jul}. The RNN and LSTM models do not capture the
intrinsic fluctuations in the counting process. The neural ODE model
is a deterministic model and cannot capture the stochasticity in the
RPA-DNA binding dynamics. Additionally, the Gaussian process can only
accurately approximate linear SDEs, which is not an appropriate form
for an SDE describing the RPA-DNA binding process.

\begin{figure*}[!htbp]
\centering
\includegraphics[width=0.9\linewidth]{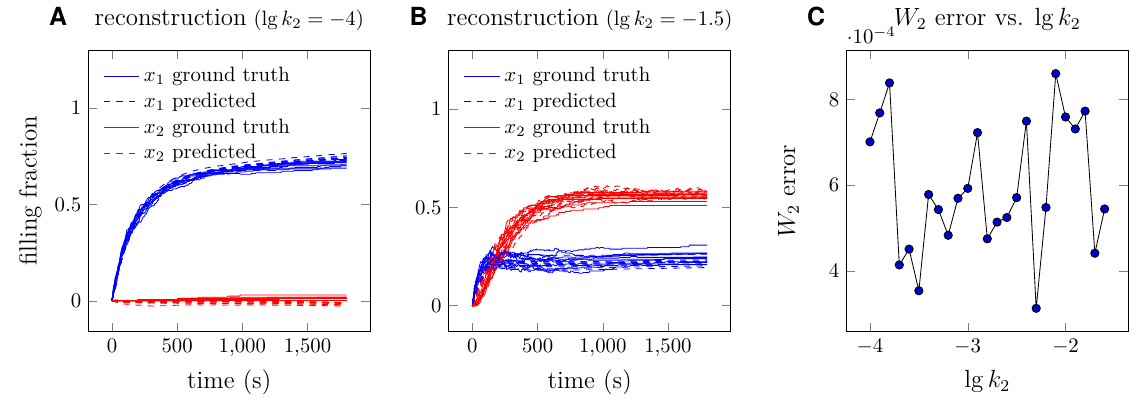}
\caption{\textbf{Reconstructed trajectories of the RPA-DNA binding
    problem.} A. Sample ground truth and reconstructed trajectories
  evaluated at $ \lg k_2 = -4 $, where we use the convention 
  that $\lg= \log_{10}$.  B. Sample ground truth and
  reconstructed parameters evaluated at $ \lg k_2 = -1.5
  $. C. Temporally decoupled squared $W_2$ distances
  (see Eq.~\eqref{eq:W_2}) between the ground truth and reconstructed
  trajectories evaluated at different $ \lg k_2 $ values. In A and B,
  blue and red trajectories represent the filling fractions of DNA by
  20nt-mode and 30nt-mode RPA, respectively.  The dashed lines
  represent the predicted trajectories, and the solid lines represent
  the ground truth.  Throughout the figure, the data are generated by
  a single neural SDE model that accepts the conversion rate $k_2$ as
  a parameter and outputs the trajectories.}
\label{fig:dna_res}
\end{figure*}
%
%

In Figs.~\ref{fig:dna_res}A,B, we plot the predicted trajectories
obtained by the trained neural SDE model for two different values $\lg
k_2 = -4 $ and $ \lg k_2 = -1.5$. Actually, for all different values
of $k_2$, trajectories generated by our END-nSDE method match well
with the ground truth trajectories on the testing set, as the
temporally decoupled squared $W_2$ loss is maintained small for all
$k_2$ (shown in Fig.~\ref{fig:dna_res}C). This demonstrates the
ability of our method to capture the dependence of the stochastic
dynamics on biochemical kinetic parameters.

\subsection{Reconstructing high-dimensional NF$\kappa$B signaling
  dynamics from simulated and experimental data}

Finally, we evaluate the effectiveness of the END-nSDE framework in
reconstructing high-dimensional NF$\kappa$B signaling dynamics under
varying noise intensities and investigate the performance of the
neural SDE method in reconstructing experimentally measured noisy
NF$\kappa$B dynamics. The procedure is divided into two parts. First,
we trained and tested our END-nSDE method on synthetic data generated
by the NF$\kappa$B SDE model Eq.~\eqref{eq:nfkb_model-3} under
different noise intensities $(\sigma_1, \sigma_2)$.  Second, we test
whether the trained END-nSDE can reproduce the experimental dynamic
trajectories.

\subsubsection{Reconstructing a 52-dimensional stochastic model for NF$\kappa$B dynamics}

\begin{figure*}
\centering
\includegraphics[width=5.4in]{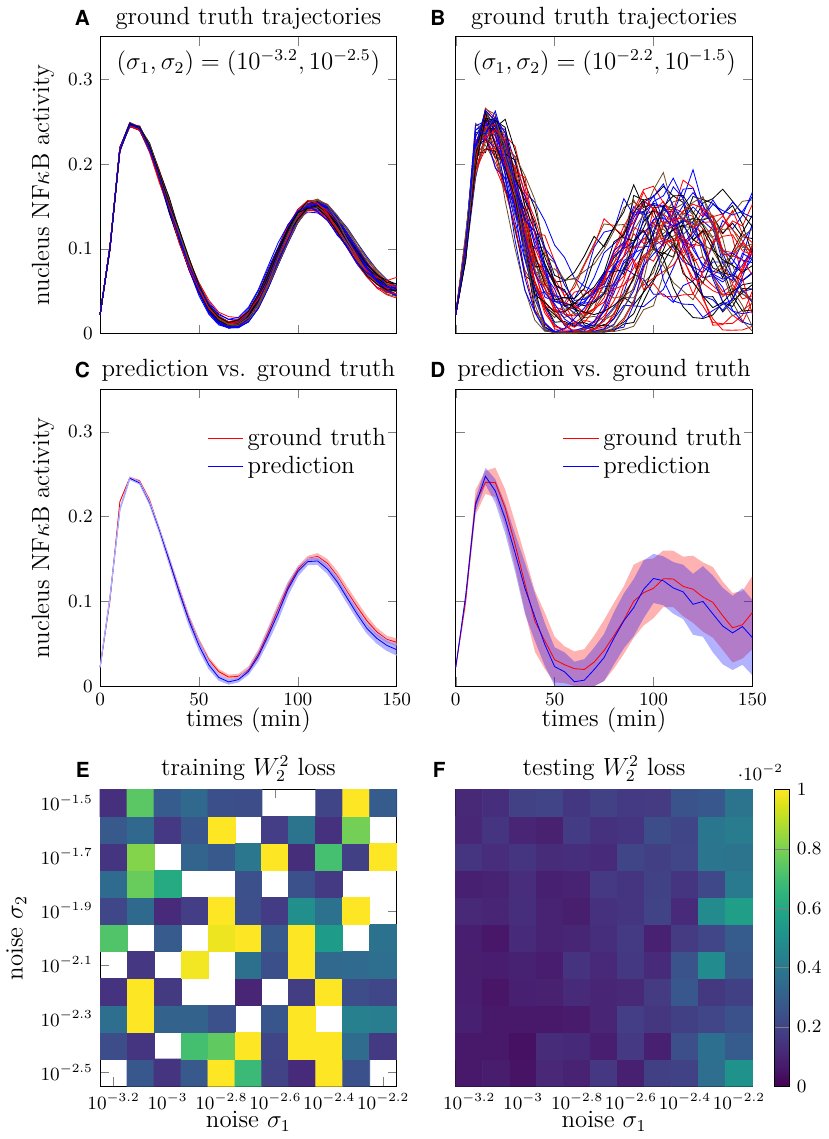}
\caption{\textbf{Reconstruction of NF$\kappa$B signaling dynamics.}
  A. Sample trajectories of nuclear NF$\kappa$B concentration as a
  function of time with $\sigma_1=10^{-3.2}$, $\sigma_2=10^{-2.5}$.
  B. Sample trajectories of nuclear NF$\kappa$B concentration as a
  function of time with $\sigma_1=10^{-2.2}$, $\sigma_2=10^{-1.5}$.
  C. Reconstructed nuclear NF$\kappa$B trajectories generated by the
  trained neural SDE versus the ground truth nuclear NF$\kappa$B
  trajectories under noise intensities $\sigma_1=10^{-3.2}$,
  $\sigma_2=10^{-2.5}$ in Eqs. \eqref{eq:nfkb_model-3}.
  D. Reconstructed nuclear NF$\kappa$B trajectories generated by the
  trained neural SDE versus the ground truth nuclear NF$\kappa$B
  trajectories under noise intensities $\sigma_1=10^{-2.2}$,
  $\sigma_2=10^{-1.5}$. E. The squared $W_{2}$ distance between the
  distributions of the predicted trajectories and ground truth
  trajectories on the training set under different noise strengths
  $(\sigma_0, \sigma_1)$. For training, we randomly selected 50\%
  sample trajectories in 80 combinations of noise strengths
  $(\sigma_1,\sigma_2)$ as the training dataset. Blank cells indicate
  that the corresponding parameter set is not included in the training
  set. F. Validation of the trained model by evaluating the squared
  $W_2$ distance between the distributions of predicted trajectories
  and ground truth trajectories on the validation set.}
\label{fig:fig1}
\end{figure*}

For training END-nSDE models, we first generate synthetic data from
the 52-dimensional SDE model of NF$\kappa$B signaling dynamics
Eqs.~\eqref{eq:nfkb_model-3} and established models
\cite{Guo2024Modeling,adelaja2021six}. The synthetic trajectories are
  generated under 121 combinations of noise intensity $(\sigma_1,
  \sigma_2)$ in Eqs.~\eqref{eq:nfkb_model-3} (see Appendix
  \ref{sec:simset_nfkb}). The resulting NF$\kappa$B trajectories vary
  depending on noise intensity, with low-intensity noise producing
  more consistent dynamics across cells (see Fig.~\ref{fig:fig1}A) and
  higher-intensity noise yielding more heterogeneous dynamics (see
  Fig.~\ref{fig:fig1}B). The simulated ground truth trajectories are
  split into training and testing datasets (see Appendix
  \ref{sec:train_nsde_nfkb} for details). Specifically, we exclude 25
  combinations of noise intensities $(\sigma_1, \sigma_2)$ from the
  training set in order to test the generalizability of the trained
  neural SDE model on noisy intensities.

Next, we trained a 52-dimensional neural SDE model using our END-nSDE
method on synthetic trajectories (see Appendix
\ref{sec:train_nsde_nfkb} for details). The loss function is based on
the $W_2$ distance between the distributions of the neural SDE
predictions in Eqs.~\eqref{eq:nfkb_model-3} and the simulated nuclear
NF$\kappa$B and I$\kappa$B$\alpha$-NF$\kappa$B complex activities
($u_5(t)$ and $u_{10}(t)$, respectively) and the corresponding
END-nSDE predictions. The remaining 50 variables of the NF$\kappa$B
system are treated as latent variables, as they are not directly
included in the loss function calculation.

Although the NF$\kappa$B dynamics vary under different noise
intensities $(\sigma_1, \sigma_2)$, the trajectories generated by our
trained neural SDE closely align with the ground truth synthetic
NF$\kappa$B dynamics under different noise intensities $(\sigma_1,
\sigma_2)$ (see Figs.~\ref{fig:fig1}C-D).  The neural SDE model
demonstrates greater accuracy in reconstructing NF$\kappa$B dynamics
when the noise in I$\kappa$B$\alpha$ transcription ($\sigma_1$) is
smaller, as evidenced by the reduced squared $W_2$ distance between
the predicted and ground-truth trajectories on both the training and
validation sets (see Figs.~\ref{fig:fig1}E-F). The temporally
decoupled squared $W_2$ loss Eq.~\eqref{eq:W_2} on the validation
set is close to that on the training set for different values of noise
intensities $(\sigma_1, \sigma_2)$. The mean squared $W_2$ distance
across all combinations of noise intensities $(\sigma_1, \sigma_2)$ is
0.0013 for the training set, and the validation set shows a mean
squared $W_2$ distance of 0.0017.

Since the loss function for this application involves only two
variables out of 52, we also tested whether the ``full''
52-dimensional NF$\kappa$B system can be effectively modeled by a
two-dimensional neural SDE. After training, we found that the reduced
model was insufficient for reconstructing the full 52-dimensional
dynamics, as it disregarded the 50 latent variables not included in
the loss function (see Fig.~\ref{fig:2d_res} in
Appendix~\ref{sec:2D_nfkb}).  This result underscores the importance
of incorporating latent variables from the system, even when they are
not explicitly included in the loss function.


\subsubsection{Reconstructing NF$\kappa$B experimental data with END-nSDE}

\begin{figure*}
\centering
\includegraphics[width=5.5in]{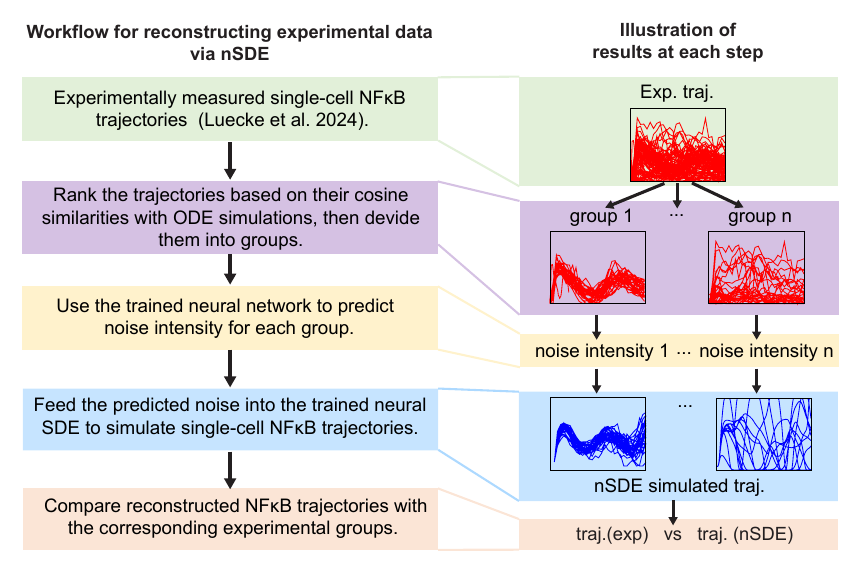}
\caption{\textbf{Workflow of reconstructing experimental data via
    END-nSDE.} Workflow for reconstructing experimental data using
  the trained parameterized nSDE and the parameter-inference neural
  network (NN). The boxes on the left outline the steps of the
  experimental data reconstruction process, while the boxes on the
  right illustrate the corresponding results at each
  step.}\label{fig:workflow_exp_vs_nSDE}
\end{figure*}

We then assess whether our proposed END-nSDE can accurately
reconstruct the experimentally measured NF$\kappa$B dynamic
trajectories. For simplicity and feasibility, we tested the END-nSDE
under the assumption that: (1) all cells share the same drift
function, and (2) cells with trajectories that deviate similarly from
their ODE predictions have the same noise intensities.  Based on these
assumptions, we developed the following workflow (see
Fig.~\ref{fig:workflow_exp_vs_nSDE}):

\begin{enumerate}
\item We used experimentally measured single-cell trajectories of
  NF$\kappa$B activity, obtained through live-cell image tracking of
  macrophages from mVenus-tagged RelA mouse with a frame frequency of
  five minutes \cite{luecke2024dynamical}. These trajectories
  correspond to the sum of nuclear I$\kappa$B$\alpha-$NF$\kappa$B and
  NF$\kappa$B in the 52D SDE model ($u_5(t)$ and $ u_{10}(t)$ in
  Eqs.~\eqref{eq:nfkb_model-3}). \\
\item The experimental dataset was divided into subgroups. Cosine
  similarity was calculated between the ODE-generated trajectory
  (representative-cell NF$\kappa$B dynamics) and experimental
  trajectories. The trajectories were then ranked and divided into
  different groups based on their cosine similarity with the ODE
  model. Experimental trajectories with higher similarity to the ODE
  trajectory are expected to exhibit smaller intrinsic fluctuations,
  corresponding to lower noise intensities (see
  Appendix \ref{sec:devide_exp_nfkb} for details).\\
\item Each group of experimental trajectories was input into the
  trained neural network (see the next paragraph for more details) to
  infer the corresponding noise intensities $(\sigma_1,
  \sigma_2)$. For simplicity, we assume that trajectories within each
  group shared the same noise intensities. \\
\item The inferred noise is then used as inputs for the trained END-nSDE to
  simulate NF$\kappa$B trajectories. \\
\item The simulated trajectories were compared with the corresponding
  experimental data to evaluate the model's performance.
  \end{enumerate}
  
To estimate noise intensities from different groups of experimentally
measured single-cell nuclear NF$\kappa$B trajectories (step (3) in the
proposed workflow), we trained another neural network to predict the
corresponding I$\kappa$B$\alpha$ transcription and NF$\kappa$B
translocation \added{noise intensities from the groups of NF$\kappa$B
  trajectories in the synthetic training data. The trained neural
  network can then be used for predicting noise intensities in the
  validation set (see Appendix~\ref{sec:infer_noise_nfkb} for
  details).}
  
\begin{figure*}
\centering
\includegraphics[width=5.8in]{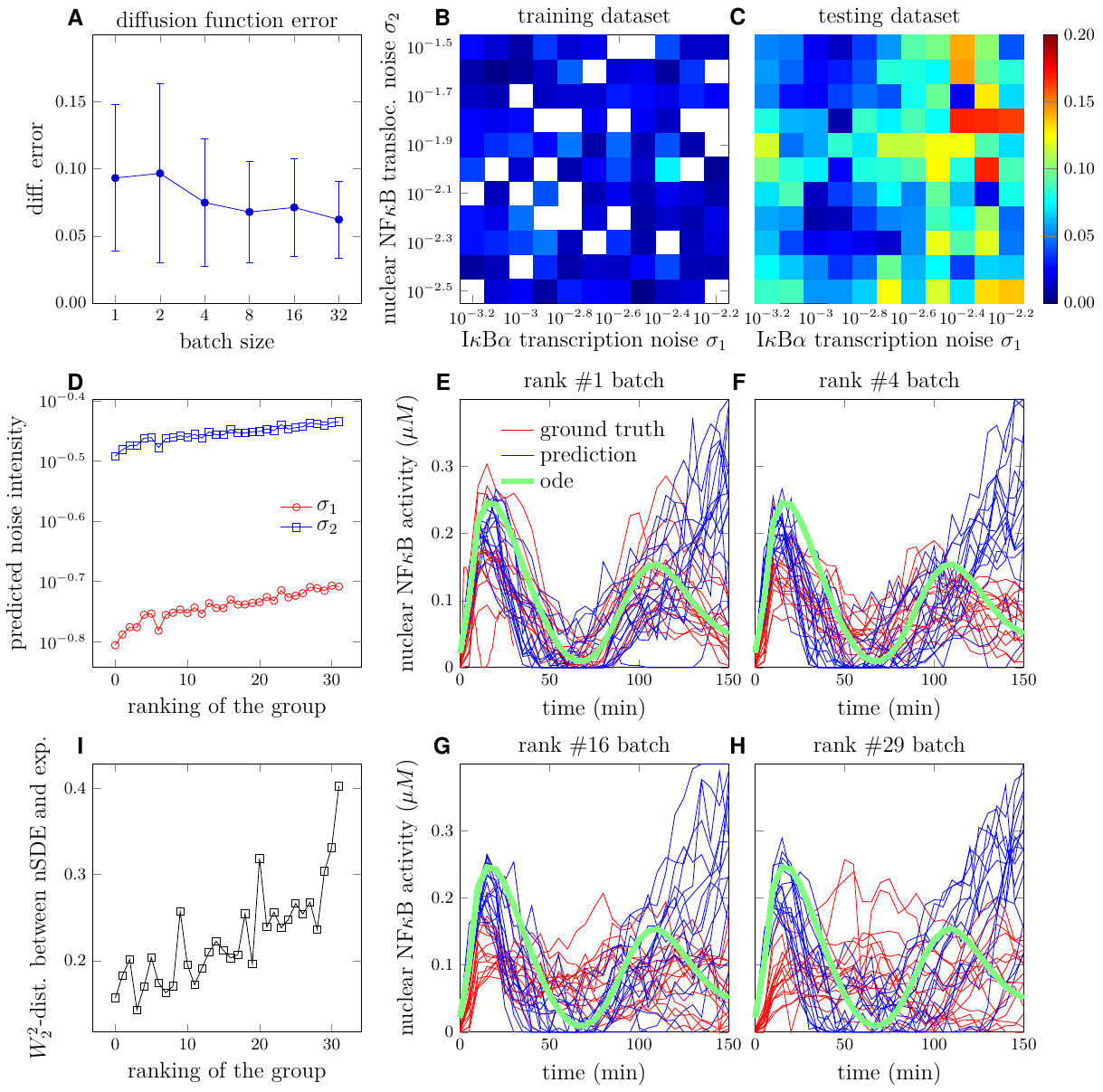}
    \caption{\textbf{Inferring intrinsic noise intensities and
        reconstructing experimental data via END-nSDE.} A. Plots
      showing the mean (solid circles) and variance (error bars) of
      the relative error in the reconstructed noise intensities
      $(\hat{\sigma}_1, \hat{\sigma}_2)$ predicted by the
      parameter-inference NN for the testing dataset, as a function of
      the group size of input trajectories. B. Heatmaps showing the
      relative error in the reconstructed noise intensities for the
      training dataset. Colored cells represent results from the
      parameter-inference NN for the training dataset, while blank
      cells indicate noise strength values not included in the
      training set. C. Heatmaps showing the relative error in the
      diffusion function for the testing dataset. D. The inferred
      intensity of I$\kappa$B$\alpha$ transcription noise ($\sigma_1$)
      and NF$\kappa$B translocation noise ($\sigma_2$) in different
      groups of experimental trajectories, plotted against the group's
      ranking in decreasing similarity with the representative ODE
      trajectory. E-H. Groups of experimental and nSDE-reconstructed
      trajectories ranked by decreasing cosine similarity: \#1 (E),
      \#4 (F), \#16 (G), \#29 (H). \deleted{The squared $W_2$-distance
        between experimental and ODE-generated trajectories are 0.077
        (E), 0.105 (F), 0.162 (G), 0.260 (H) while} The squared
      $W_2$-distance between experimental and SDE-generated
      trajectories are 0.157 (E), 0.143 (F), 0.212 (G), 0.236 (H). The
      inferred noises are $(10^{-0.49},10^{ -0.81})$ (E),
      $(10^{-0.47},10^{-0.78})$ (F), $(10^{-0.46},10^{-0.74})$ (G),
      $(10^{-0.44},10^{-0.71})$ (H). I. The temporally decoupled
      squared $W_2$ distance between reconstructed trajectories
      generated by the trained END-nSDE and groups of experimental
      trajectories, ordered according to decreasing similarity with
      the representative ODE trajectory.}
    \label{fig:exp_vs_nSDE}
\end{figure*}

Assessing the impact of group size (number of trajectories) on noise
intensity prediction performance, we found that taking a group size of
at least two leads to a relative error of around 0.1 (see
Fig.~\ref{fig:exp_vs_nSDE}A). Given the high heterogeneity present in
experimental data, we took a group size of 32 as the input into the
neural network. Under this group size, the relative errors in the
predicted noise intensities were 0.021 on the training set and 0.062
on the testing set (see Figs.~\ref{fig:exp_vs_nSDE}B-C).

Using the trained neural network, we inferred noise intensities for
the experimental data, which were grouped based on their cosine
similarities with the representative-cell trajectory (deterministic
ODE) with a group size of 32. The predicted noise intensities on the
experiment data set are larger than the noise intensities on the
training set, and the underlying reason could be extrinsic noise which
is not taken into account interferes with the inference of noise
intensity.  The transcription noise of I$\kappa$B$\alpha$ is predicted
to be within the range of $[10^{-0.81}, 10^{-0.71}]$
(see Fig.~\ref{fig:exp_vs_nSDE}D). In addition, the inferred noise for
NF$\kappa$B translocation fell within $[10^{-0.49}, 10^{-0.43}]$
(see Fig.~\ref{fig:exp_vs_nSDE}D). These inferred noise intensities were
then used as inputs to the END-nSDE to simulate NF$\kappa$B
trajectories.

We compare the reconstructed NF$\kappa$B trajectories generated by the
trained neural SDE model with the experimentally measured NF$\kappa$B
trajectories (see Figs.~\ref{fig:exp_vs_nSDE}E-I). The trajectories
generated using our END-nSDE method successfully reproduce the
experimental dynamics for the majority of time points for the top 50\%
of cell subgroups most correlated with the representative-cell ODE
model (see Figs.~\ref{fig:exp_vs_nSDE}E-G, Figs.~\ref{fig:exp_vs_nSDE}I).

For the top-ranked subgroups (\#1 to \#16), the heterogeneous
nSDE-reconstructed dynamics align well with the experimental data for
the first 100 minutes. The predicted trajectories deviate more from
ground truth trajectories observed in experiments after 100 minutes
possibly due to error accumulation and errors in the predicted noise
intensity.
For experimental subgroups that significantly deviate from the
representative-cell ODE model, the END-nSDE struggles to fully capture
the heterogeneous trajectories. This limitation likely arises from the
assumption that all cells in a group share the same underlying
dynamics, whereas in reality, substantial cellular differences in
underlying dynamics exist due to heterogeneity in the drift term, an
aspect not accounted for in END-nSDE due to the high computational
cost.

\deleted{Nonetheless, the average squared $W_2$-distance between the
  experimental data and the ODE trajectory is 0.18 among all groups of
  experimental data, whereas the average squared $W_2$-distance
  between the reconstructed SDEs and the experimental data is
  0.23. This demonstrates that our END-nSDE can train a neural SDE
  model that better captures the noisy NF$\kappa$B signaling dynamics
  compared to the previous deterministic ODE model, which is unable to
  account for fluctuations in the dynamics.}

\added{Nonetheless, our END-nSDE can partially reconstruct
  experimental datasets and has the potential to fully capture
  experimental dynamics. With sufficient computational resources, our
  proposed workflow can also incorporate extrinsic noise in the
  cellular dynamics drift terms, allowing for further discrimination
  of experimental trajectories.}

Overall, we have demonstrated that
reconstructing noisy experimental trajectories can be accomplished by
(i) inferring noise intensities from noisy trajectories grouped by
different noise levels and (ii) using END-nSDE to reconstruct the
experimental data based on the inferred noise.

\section{Discussion} 

In this work, we used a $W_{2}$-distance to develop an END-nSDE
reconstruction method that takes into account extrinsic noise in gene
expression dynamic as observed across various biophysical and
biochemical processes such as circadian rhythms, RPA-DNA binding, and
NF$\kappa$B translocation.  We first demonstrated that our END-nSDE
method can successfully reconstruct a minimal noise-driven fluctuating
SDE characterizing the circadian rhythm, showcasing its effectiveness
in reconstructing SDE models that contain both intrinsic and extrinsic
noise. Next, we used our END-nSDE method to learn a surrogate
extrinsic-noise-driven neural SDE, which approximates the RPA-DNA
binding process. Molecular binding processes are usually modeled by a
Markov counting process and simulated using Monte-Carlo-type
stochastic simulation algorithms (SSAs) \cite{Gillespie1977dec}.  Our
END-nSDE reconstruction approach can effectively reconstruct the
stochastic dynamics of the RPA-ssDNA binding process while also taking
into account extrinsic noise (heterogeneity in biological parameters
among different cells). Our END-nSDE method outperforms several
benchmark methods such as LSTMs, RNNs, neural ODEs, and Gaussian
processes.

Finally, we applied our methodology to analyze NF$\kappa$B
trajectories collected from over a thousand cells. Not only did the
neural SDE model trained on the synthetic dataset perform well on the
validation set, but it also partially recapitulated experimental
trajectories of NF$\kappa$B abundances, particularly for subgroups
with dynamics similar to the representative cell. These results
underscore the potential of neural SDEs in modeling and understanding
the role of intrinsic noise in complex cellular signaling systems
\cite{rao2002control,arias2006filtering, eling2019challenges}.

When the experimental trajectories were divided into subgroups, we
assumed that all cells across different groups shared the same drift
function (as in the representative ODE) and cells within each group
shared the same diffusion term. We found that subgroups with dynamics
more closely aligned with the deterministic ODE model resulted in
better reconstructions. In contrast, for experimental trajectories
that deviated significantly from the representative ODE model, their
underlying dynamics may differ from those defined by the
representative cell's ODE. Therefore, the assumption that a group
shares the same drift function as the representative cell ODE holds
only when the trajectories closely resemble the ODE. Incorporating
noise into the drift term for training the neural SDE could
potentially address this issue. We did not consider this approach due
to the high computational cost required for training.

Applying our method to high-dimensional synthetic NF$\kappa$B
datasets, we showed the importance of incorporating latent
variables. This necessity arises because the ground-truth dynamics of
the measured quantities (nuclear NF$\kappa$B) are not self-closed and
inherently depend on additional variables. Consequently, the
52-dimensional SDE reconstruction requires more variables than just
the ``observed'' dynamics of nuclear NF$\kappa$B. In this example, the
remaining 50 variables in the nSDE were treated as latent variables,
even though they were not explicitly included in the loss function.



There are several promising directions for future research. First, it
is desirable to identify methods to extract an explicit and
interpretable version of the learned neural network SDEs. One might
employ a polynomial model for the drift and diffusion functions in the
SDE similar to the one suggested in
Ref.~\cite{fronk2023interpretable}. Such an explicit representation
may facilitate linking neural reconstruction methods to established
mechanistic models. The importance of latent variables should also be
analyzed to understand if and how they affect the reconstruction of
high-dimensional models. The $W_{2}$-distance loss landscapes in the
examples examined in this study should also be examined in more
depth. Prior research~\cite{bottcher2022near,bottcher2024visualizing}
has highlighted the importance of studying loss landscapes to
characterize and potentially enhance the generalization capabilities
of neural networks across various tasks. Finally, neural SDEs can
serve as surrogate models for complex biomedical
dynamics~\cite{fonseca2024metamodeling,bottcher2024control}. Combining
such surrogate models with neural control
functions~\cite{asikis2022neural,bottcher2022near,bottcher2022ai} can
be useful for tackling complex biomedical control problems.

\section*{Acknowledgments}
LB acknowledges financial support from hessian. AI and the ARO through
grant W911NF-23-1-0129. TC acknowledges inspiring discussions at the
``Statistical Physics and Adaptive Immunity'' program the Aspen Center
for Physics, which is supported by National Science Foundation grant
PHY-2210452. We acknowledge Stefanie Luecke providing the experimental
datasets for NF$\kappa$B dynamics.

\bibliography{ref}

\appendix

\setcounter{figure}{0}
\renewcommand{\thefigure}{S\arabic{figure}}
\section{Hyperparameters and conditions used in the neural
  SDE model}\label{sec:sdeap} In this section, we provide the
hyperparameters in the neural SDE models as well as the training
details in Table ~\ref{tab:all-sde}.

\begin{table*}[!htbp] 
    \centering
    \caption{The hyperparameters for training the neural SDE model of
      each example.}
    \begin{tabular}{l|ccc}
        \hline
         & Example 1 & Example 2 & Example 3 \\
        \hline
        Gradient descent method & Adam & Adam & Adam\\
       Learning rate & 0.002 & 0.002 & 0.002  \\
        Weight decay & 0.005 & 0.005 & 0.005  \\
        Activation function & ReLU & ReLU & ReLU\\
        \# of epochs & 500 & 2000 & 2000 \\
        \# of hidden layers in $\hat{f}$ &3&2&2\\
         \# of neurons in hidden layer in $\hat{f}$ & 100 & 200 & 400\\
                 \# of hidden layers in $\hat{\sigma}$ &3&2&2\\
         \# of neurons in each hidden layer in $\hat{\sigma}$ & 100 & 200 & 400\\
$\Delta t$ & 0.1 (s) & $\frac{1}{720}$ (hr) & $\frac{1}{12}$ (hr)\\
$T$ & 11 & 361 & 31\\
\hline
        \hline
    \end{tabular}\label{tab:all-sde}
\end{table*}

\section{RPA dynamic binding to long ssDNA can be simulated by a
generalized random sequential adsorption (RSA) model}
\label{sec:dna_binding}
To elucidate the biophysical mechanism of replication protein A (RPA)
binding dynamics to long single-stranded DNA (ssDNA), we developed a
continuous-time discrete Markov chain model (see Fig.~\ref{fig:4ab}). The
Random Sequential Adsorption (RSA) model in a one-dimensional (1D)
finite-length context effectively represents the process of protein
binding to DNA, capturing the key property that each nucleotide (nt)
of ssDNA cannot be occupied by more than one protein molecule. This
unique characteristic leads to incomplete occupation even with protein
oversaturation, distinguishing DNA-relevant reactions from those
described by the mass action law. To reveal finer structures such as
gap distribution, we implemented an exact stochastic sampling
approach.

We adapted the model based on current knowledge of RPA binding modes,
incorporating multiple binding modes and volume exclusion effects. RPA
has two binding modes: a 20-nt mode (partial binding mode, PBM) and a
30-nt mode (full-length binding mode, FLBM) (see Fig.~\ref{fig:4ab}). RPA
initially binds to a 20-nt ssDNA with a rate of \( k_1 \) and
dissociates at a rate of \( k_{-1} \). This 20-nt mode assumes
constant \( k_1 \). One scenario involves
DBD-A, DBD-B, and DBD-C binding to ssDNA, with subsequent DBD-D
binding leading to the 30-nt mode. In the 20-nt mode, DBD-D binds an
extra 10-nt ssDNA with a rate of \( k_2 \) and dissociates at \(
k_{-2} \), forming the 30-nt mode (FLBM). RPA always aligns in the
same direction along DNA.

The multiple binding modes result in interesting kinetic features like
facilitated exchange and desorption. In this model, the ssDNA fragment
state is represented by a vector of length \( L \) with each component
taking values in \(\{0,1\}\), where 0 indicates an unoccupied
nucleotide and 1 indicates an occupied one. Each RPA initially
occupies \(\ell = 20\) nts and can further occupy \(\Delta \ell = 10\)
nts if the local state allows, moving in the 3’ direction. Each DNA
site can be occupied by only one RPA molecule, altering the RSA
model's available reactions.  Each consecutive unoccupied segment of
length \(\ell\) recruits one RPA at the rate of \( k_1 \), with the
total binding rate given by:
\[
v_1 = k_1 \sum_{j=1}^{L-\ell+1} \delta_{\mathbf{0}^{\times \ell}} (\operatorname{state}[j, j+1, \ldots, j+\ell-1]),
\]
where \(\delta_{a}(b)\) is the Kronecker delta function,
\(\mathbf{0}^{\times \ell}\) is a zero vector of length \(\ell\), and
\(\operatorname{state}[j, \ldots, j+\ell-1]\) is the ssDNA state
vector.

To occupy another 10 nts, we assign\added{ed} a rate parameter \( k_2 \) and
calculate the overall rate by:
\[
v_2 = k_2 \sum_{q_j} \delta_{\mathbf{0}^{\times \Delta \ell}}
\left( \operatorname{state}[q_j+\ell, \ldots, q_j+\ell+\Delta \ell-1] \right),
\]
where \( q_j \) represents the leftmost position of each bound RPA in
the 20-nt mode. For unbinding, the 30-nt mode reopens to the 20-nt
mode at rate \( k_{-2} \), and the 20-nt mode desorbs at rate \(
k_{-1} \).  The overall rates are:
\[
\begin{aligned}
& v_{-2} = k_{-2} \# \{\text{FLBM RPA}\}, \\
& v_{-1} = k_{-1} \# \{\text{PBM RPA}\}.
\end{aligned}
\]
The total possible reaction rate is:
\[
v_{\mathrm{tot}} = v_1 + v_{-1} + v_2 + v_{-2}.
\]

Reactions occur stochastically according to exponentially distributed
waiting times with parameter \( v_{\mathrm{tot}} \). The waiting time
\( \delta t \) follows the exponential distribution with a probability
density function:
\[
\operatorname{pdf}(\delta t = t) = v_{\mathrm{tot}} e^{-v_{\mathrm{tot}} t}.
\]
After each reaction, the DNA state updates, and the possible reactions
are re-evaluated. The Gillespie algorithm was used to sample the
trajectories of this stochastic model. We used Julia to perform exact
stochastic simulations of all known RPA-ssDNA interactions, and codes
are available at (\url{https://github.com/hsianktin/RPA_model}).

\section{Implementation of benchmarks}

In this section, we introduce the benchmark methods; hyperparameters
and settings for training of all methods used in this paper are shown
in Table~\ref{tab:tab2}.

\subsection{RNN networks}
\label{sec:rnn_}
\begin{figure}[htb]
\centering
\includegraphics[width=\linewidth]{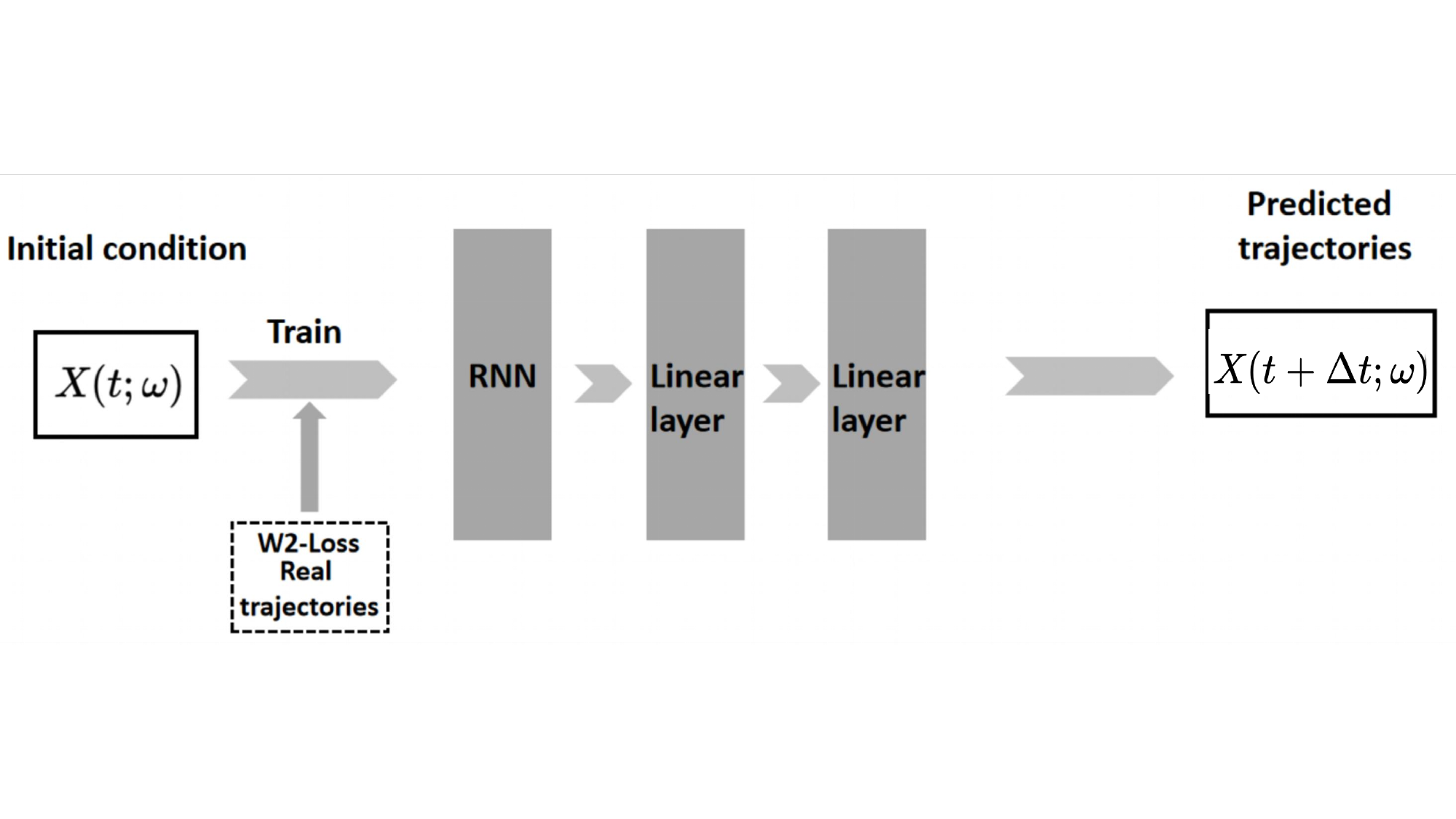}
\caption{The structure of the RNN model used. The RNN layer in this
  figure can be built using \texttt{torch.nn.RNN}. The structure of
  RNN in the figure can be found in Ref.~\cite{pascanu2013construct}.}
\label{fig:rnn}
\end{figure}

Recurrent neural networks (RNN) are often used for language processing,
but they can also be used to analyze time series data with temporal
correlations. For reconstructing the RPA-DNA binding dynamics, we used
the RNN model in Ref.~\cite{computation11020039}. A neural network
that contains two layers of RNN and two linear layers was used to model
RPA's dynamic binding with single-stranded DNA. All layers of RNN are
built using \texttt{torch.nn.Module} package. Hyperparameters in the
neural network is initialized by default. The parameters of this
neural network are initialized by default. At each time step,
$\bm{X}(t;\omega) \in \mathbb{R}^{d}$ representing the RPA dynamics at
this time point is inputted in Example~\ref{ex_dna}, and the dynamics
at the next time point: $\bm{X}(t+\Delta t;\omega)$ is outputted as the
prediction.
The RNN is trained by optimizing the loss function
Eq.~\eqref{eq:W_2}. The pipeline of the model could be seen in
Fig.~\ref{fig:rnn}.

\subsection{LSTM networks}
\label{sec:lstm}
\begin{figure}[htb]
\centering
\includegraphics[width=\linewidth]{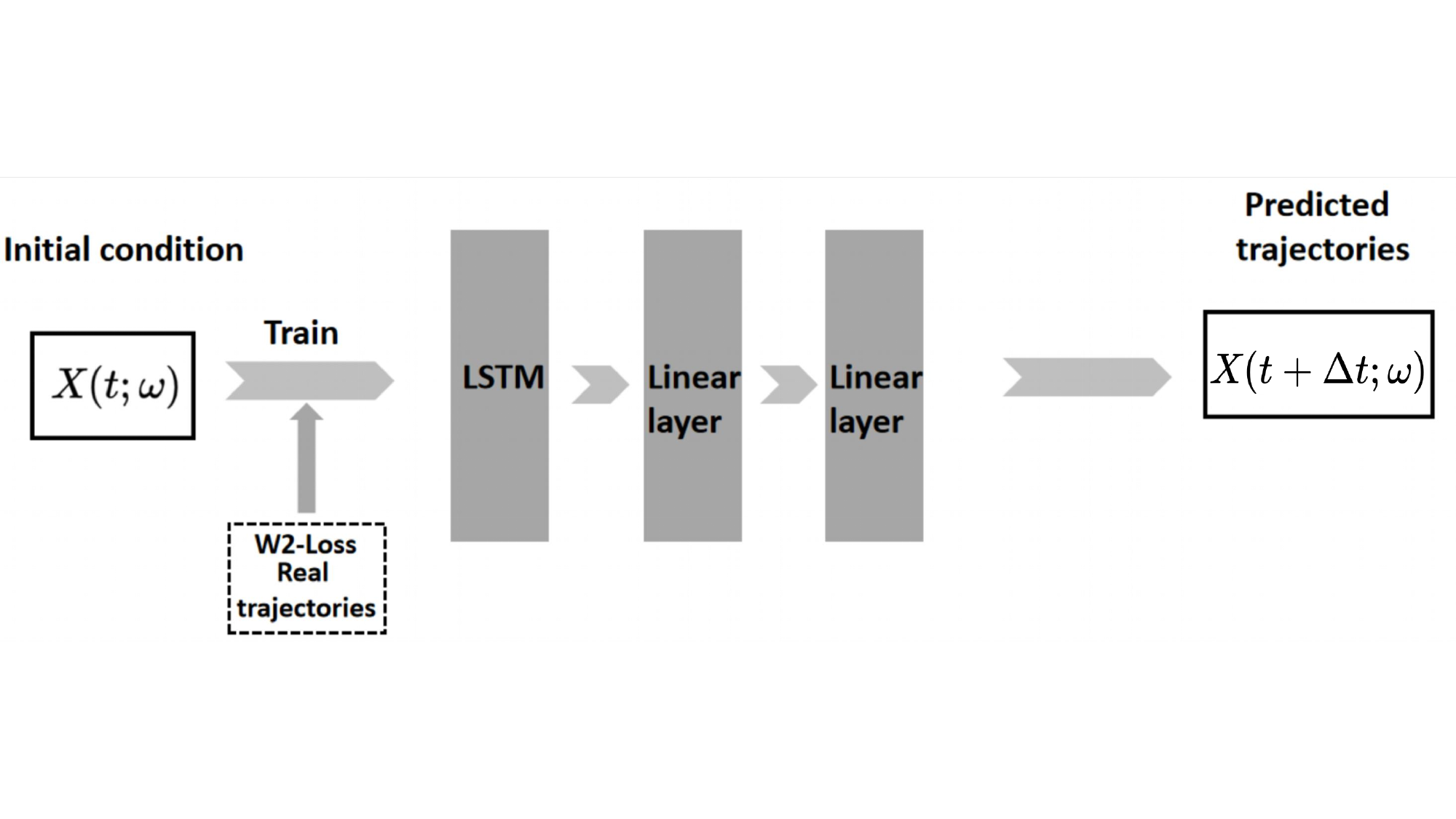}
\caption{The structure of the LSTM model used. The LSTM layer in this
  figure can be built using \texttt{torch.nn.LSTM}. The structure of
  LSTM in figure can be found in \cite{6795963}.} \label{fig:lstm}
\end{figure}

Long Short-Term Memory (LSTM) networks \cite{hochreiter1997long}, a
class of variants of Recurrent Neural Networks (RNNs), have been
widely used in modeling sequential data such as time series and
natural language.  We used the LSTM network model proposed in
Ref.~\cite{wang2019new} implemented through the
\texttt{torch.nn.Module} package for reconstructing the RPA-DNA
dynamics. Hyperparameters in the neural network is initialized by
default.  At each time step, the $\bm{X}(t;\omega) \in \mathbb{R}^{d}$
discussed above is inputted in Example~\ref{ex_dna}, and the state at
the next time point: $\bm{X}(t+\Delta t;\omega)$ is outputted as the
prediction. Then, the gradients of the loss function
Eq.~\eqref{eq:W_2} are calculated to update the parameters in the
LSTM model. The input size of the two layers of LSTM is 2 and the
hidden size is 4. The input size and output size of the two linear
layers are 8,4 and 4,2 respectively.  The LSTM is trained by
optimizing the loss function Eq.~\eqref{eq:W_2}. The pipeline of
the model is shown in Fig.~\ref{fig:lstm}.

\subsection{Gaussian Process}
\label{sec:GP}
The Gaussian Process (GP) \cite{conf/ac/Rasmussen03} is also widely
used in modeling time-series data; it is implemented using the
\texttt{gaussianprocess.GaussianProcessRegressor} package in
Python. We used radial basis functions as kernel functions (using
sklearn.gaussianprocess.kernels package), and the hyperparameters of
$\alpha$ and n-restarts-optimizer are 0.3 and 5, respectively. When
training the GP model, we inputted the trajectories of all time points
in the training set to the GP for fitting. Then, we inputted the
trajectory of the current time point in the testing set, and the model
predicts the trajectory of the next time point based on the
kernel. The detailed structure of the GP model we used is the same as
that in Ref.~\cite{williams1995gaussian}.

\begin{figure}[htb]
\centering
\includegraphics[width=\linewidth]{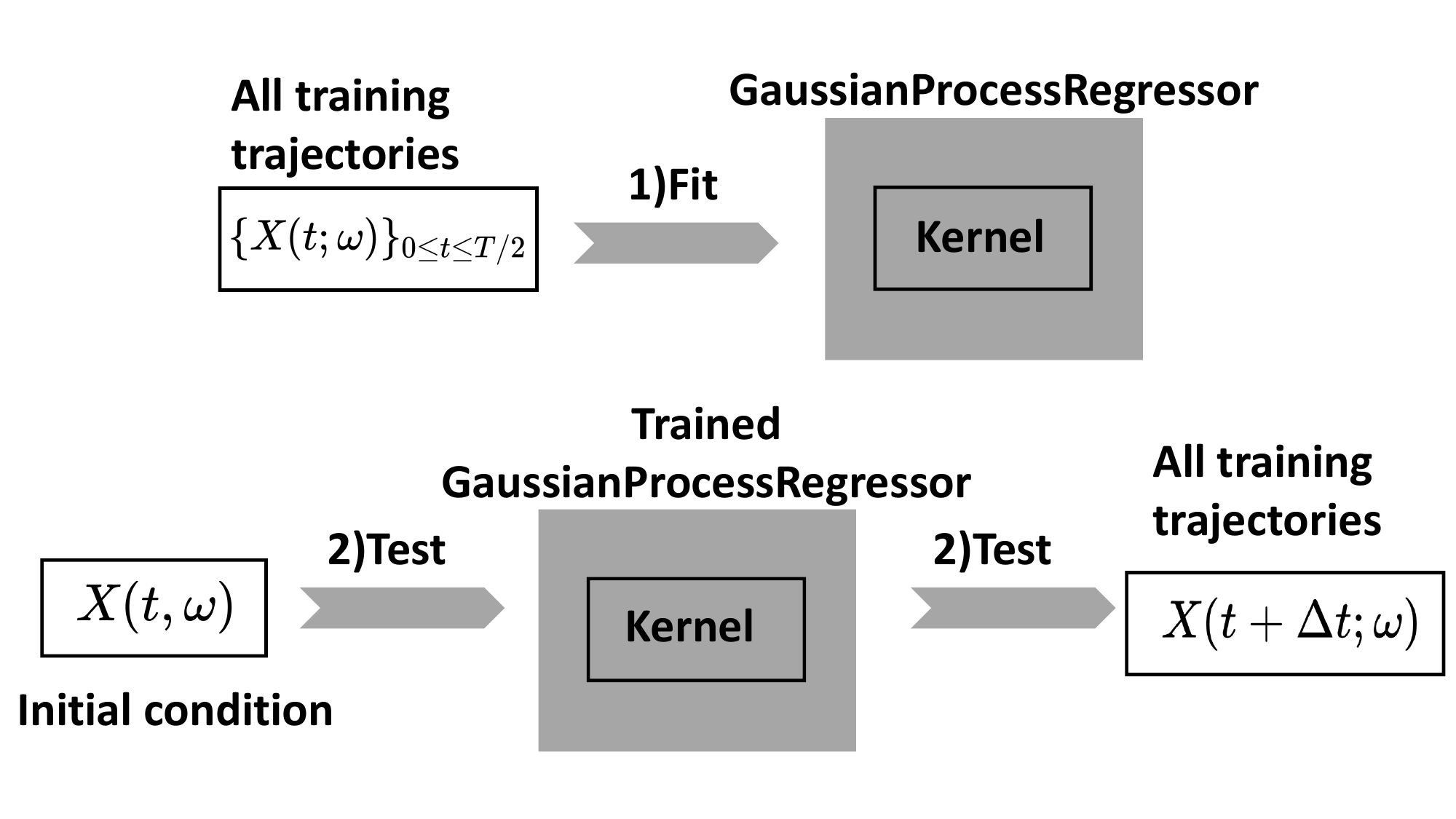}
\caption{The structure of the GP model used in this
  work.} \label{fig:gp}
\end{figure}

\begin{table*}[!htbp] 
\caption{\textbf{Hyperparameters of benchmark methods}}
\centering
\begin{tabular}{ccc}
\hline   Model & Implementation \\
\hline   RNN & one RNN layer plus four MLP layers  \\
             &  width of RNN layer: 256\\
             &  learning rate: $5 \times 10^{-4}$\\
             & learning epoch: 2000\\
             & Optimization method: Adam\\
\hline   LSTM & one LSTM layer plus four MLP layers  \\
             &  width of LSTM layer: 128\\
             &  learning rate: $5 \times 10^{-4}$\\
             & learning epoch: 3000\\
             & Optimization method: Adam\\
\hline
\end{tabular}
\label{tab:tab2}
\end{table*}

\section{Generating the simulated NF$\kappa$B dataset}
\label{sec:simset_nfkb}

To evaluate the performance of our proposed END-nSDE method in
reconstructing dynamics of NF$\kappa$B signaling, we generated
training trajectories by simulating a previously developed
\added{52-dimensional model \cite{Guo2024Modeling,adelaja2021six} for
  the NF$\kappa$B signaling network under 100ng/mL TNF stimulation
  (see Eqs.~\eqref{eq:nfkb_model-3} for the SDEs).}  The synthetic
trajectories were generated using 100 sets of noise intensities. We
set $\sigma_1\in[10^{-3.2},10^{-2.2}],
\sigma_2\in[10^{-2.5},10^{-1.5}]$ in Eqs.~\eqref{eq:nfkb_model-3} and
use 100 combinations of $(\sigma_1, \sigma_2)$: $\sigma_1 \in \{
10^{-3.2+i\delta_1}, i=0,\dots,9\}, \delta_1=0.1$ and $\sigma_2 \in \{
10^{-2.5+j\delta_2}, j=0,\dots,9\}, \delta_2=0.1$ respectively. Other
parameters were fixed constants. The parameter values in
Eqs.~\eqref{eq:nfkb_model-3} are listed in
Table~\ref{tab:nfkb-para}. The parameters for the remaining equations
are the same as those in Refs.~\cite{Guo2024Modeling,adelaja2021six}.

The corresponding SDEs were simulated using the 'SDEProblem' function
from the 'DifferentialEquations' package in Julia. Simulations were
conducted from 0 minutes (stimulus application time) to 150 minutes,
and the state was recorded at every 5 minute intervals. Initial values
were set to the steady-state solutions of the ordinary differential
equations (ODEs), which were obtained using the ode15s function in
MATLAB.

\begin{table*}
\centering
\caption{Parameter values for the NF$\kappa$B model.}
\begin{tabular}{l l l}
\hline
\textbf{Parameter} &  \textbf{Value} & \textbf{description} \\
\hline
$k_{\text{basal}}$   & $5 \times 10^{-7}$ & basal I$\kappa$B$\alpha$ mRNA synthesis \\
$k_{\text{max}}$   & $6 \times 10^{-5}$ & maximal rate of I$\kappa$B$\alpha$ mRNA synthesis induced by NF$\kappa$B \\
$n_{\text{NF}\kappa\text{B}}$ & 2.938 & Hill coefficient for mRNA syn \\
$K_{\text{NF}\kappa\text{B}}$ & 0.1775 & EC50 for mRNA syn\\
$k_{\text{deg}}$ & 0.33 & degradation rate of I$\kappa$B$\alpha$ mRNA \\
$k_{\text{imp}}$ & 0.6 & import rate of NF$\kappa$B \\
$k_{\text{a-I}\kappa\text{B-NF}\kappa\text{B}}$ & 200 & association rate of NF$\kappa$B and  I$\kappa$B$\alpha$\\
$k_{\text{deg-NF}\kappa\text{B}}$ & 0 & degradation rate of NF$\kappa$B \\
$k_{\text{exp}}$ & 0.042 & export rate of NF$\kappa$B \\
$v$ & 3.5 & Volume ratio: cytoplasmic volume/nuclear volume \\
$k_{\rm{d\text{-}I\kappa B{\text-}NF\kappa B}}$ & 0.008 & dissociation rate of NF$\kappa$B-I$\kappa$B$\alpha$ \\
$k_{\text{phos}}$ & 2 & Phosphorylation/degradation of complexed I$\kappa$B$\alpha$\\
\hline
\end{tabular}
\label{tab:nfkb-para}
\end{table*}

\section{Training a neural SDE using simulated NF$\kappa$B trajectories}
\label{sec:train_nsde_nfkb}
We partitioned the simulated ground-truth trajectories into training
and validation sets as follows: \added{50\% of each of 96 sets of
  trajectories (out of a total of 121) associated with different noise
  intensities were used for training, while the remaining simulated
  ground-truth trajectories were used as the validation dataset.}
Since we could only observe NF$\kappa$B activity, we defined our loss
function (see Eq.~\eqref{eq:W_2}) to focus solely on the nuclear
NF$\kappa$B and the I$\kappa$B$\alpha$-NF$\kappa$B nuclear complex. In
the loss function Eq.~\eqref{eq:W_2}, $\omega\equiv(\sigma_1,
\sigma_2)$ denoted two noise intensities (see
Eqs.~\eqref{eq:nfkb_model-3}).  $\mu(\omega)$ and $\hat{\mu}(\omega)$
are the distributions of $\bm{X}(t;\omega)$ and
$\hat{\bm{X}}(t;\omega)$, respectively.  $\bm{X}(t;\omega)$
represented the values of $(u_5, u_{10})$ in
Eqs.~\eqref{eq:nfkb_model-3} at time $t$. The other 50 variables were
not included in the calculation of the loss function.

\section{Reconstructing I$\kappa$B$\alpha$NF$\kappa$Bn and NF$\kappa$Bn in
  NF$\kappa$B signaling as a 2D SDE model}
\label{sec:2D_nfkb}
\begin{figure*}
\centering
\includegraphics[width=7in]{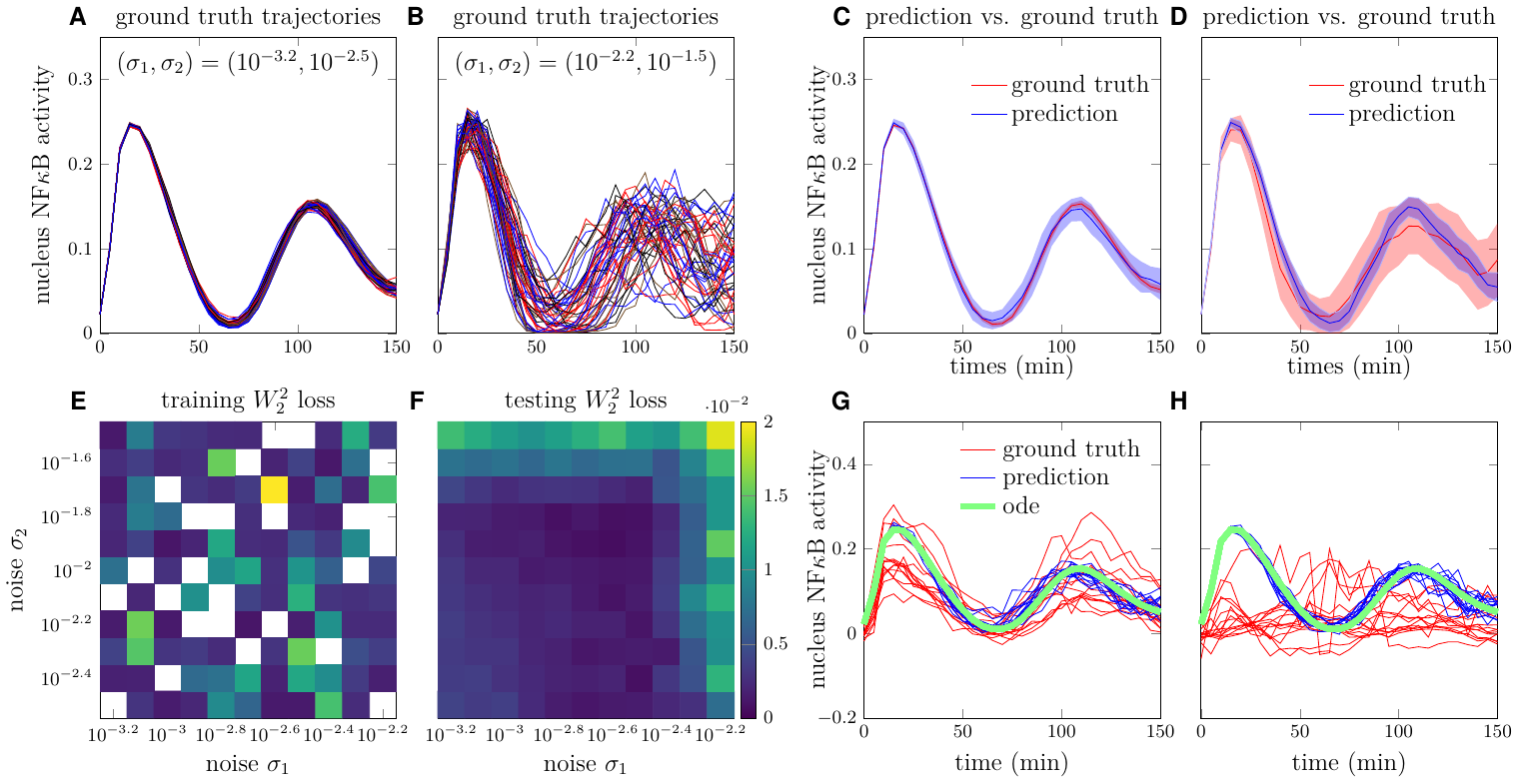}
\caption{\textbf{Reconstructed NF$\kappa$B dynamics using a
    two-dimensional nSDE.}  A. Trajectories of nuclear NF$\kappa$B
  concentration over time in the synthetic dataset with noise
  intensities $\sigma_1=10^{-3.2}$,
  $\sigma_2=10^{-2.5}$. B. trajectories of nuclear NF$\kappa$B
  concentration over time in the synthetic dataset with noise
  intensities $\sigma_1=10^{-2.2}$,
  $\sigma_2=10^{-1.5}$. C. Comparison of NF$\kappa$Bn trajectories
  predicted by the neural SDE with the ground truth trajectories under
  noise intensities $\sigma_1=10^{-3.2}$, $\sigma_2=10^{-2.5}$. D.
  Comparison of NF$\kappa$Bn predicted by the neural SDE with the
  ground truth trajectories under noise intensities
  $\sigma_1=10^{-2.2}$, $\sigma_2=10^{-1.5}$. E. The squared $W_{2}$
  distance between distributions of predicted trajectories and ground
  truth trajectories in the training set under different noise
  strengths $\sigma_0, \sigma_1$.  Empty cells indicate that the
  corresponding parameter set is not included in the training set.  F.
  The squared $W_2$ distance between distributions of the predicted
  trajectories and the ground truth trajectories on the testing
  set. \added{G.-H. After grouping the experimental trajectories by
    cosine similarity to the ODE reference trajectory (shown in
    green), the trained neural network estimated the noise
    ($\sigma_1$, $\sigma_2$) of each group of experimental
    trajectories (shown in red). Then, the estimated noise was fed
    into the nSDE to generate a group of (reconstructed) trajectories
    (shown in blue). The highest- and lowest-ranked similarity groups
    (\#1 and \#31, see Fig.~\ref{fig:exp_vs_nSDE}D, E, I) are shown in
    G. and H., respectively.}}
\label{fig:2d_res}
\end{figure*}

To investigate the importance of latent variables that are not
included in the loss function, instead of reconstructing the 52D
surrogate SDE model for the NF$\kappa$B signaling dynamics, we
attempted to directly reconstruct the dynamics of the nuclear complex
I$\kappa$B$\alpha-$NF$\kappa$B and nuclear NF$\kappa$B using the 2D
SDE
\begin{equation}
\begin{aligned}
  \d \hat{u}_5 = f_5(\hat{u}_5, \hat{u}_{10}, t)\text{d}t
  + \sigma_1(\hat{u}_5, \hat{u}_{10}, t)\text{d}B_{1,t},\\
    \d \hat{u}_{10} = f_{10}(\hat{u}_5, \hat{u}_{10}, t)\text{d}t
    + \sigma_2(\hat{u}_5, \hat{u}_{10}, t)\text{d}B_{2,t}
\end{aligned}
\label{eq:2D_SDE}
\end{equation}
to approximate the two SDEs in NF$\kappa$B signaling model
\cite{Guo2024Modeling,adelaja2021six} that describes nuclear
I$\kappa$B$\alpha$-NF$\kappa$B complex and
NF$\kappa$B. \added{Simulations and reconstruction of this model are
  shown in Fig.~\ref{fig:2d_res}.}

Using a 2D SDE model to reconstruct the NF$\kappa$B signaling dynamics
could not accurately reconstruct the noisy dynamics of nuclear
I$\kappa$B$\alpha$-NF$\kappa$B and NF$\kappa$B. As shown in
Figs. \ref{fig:2d_res}G-H, for certain noise strengths $(\sigma_0,
\sigma_1)$, both the training and testing losses (the temporally
decoupled squared $W_{2}$ distance Eq.~\eqref{loss}) are large
compared to those obtained from the direct reconstruction of the full
52-dimensional model. Furthermore, when providing the inferred noise
strengths from experimentally observed trajectories (the same as in
Figs.~\ref{fig:fig1}A, B), the reconstructed 2D SDE fails to generate
predicted trajectories $(\hat{u}_5, \hat{u}_{10})$ that align well
with experimental data. Thus, it is necessary to retain the remaining
50 variables in the model, although they are not directly used in the
calculation of the loss function.


\section{Dividing experimental NF$\kappa$B trajectories into subgroups}
\label{sec:devide_exp_nfkb}
We divided the experimentally observed trajectories into 32 groups,
each consisting of 32 trajectories. The experimental data were divided
into different groups based on each trajectory's correlation with a
ODE-model-based deterministic
trajectory~\cite{adelaja2021,Guo2024Modeling}
\begin{equation}~\label{eq:simi}
  \text{corr}(\mathbf{v}_{j}, \mathbf{v}_{\text{ODE}})
  = \frac{\sum_{i=1}^{n} v_{j}(t_i) v_{\text{ODE}}(t_i)}{\sqrt{\sum_{i=1}^{n} v_{j}^2(t_i)} \cdot \sqrt{\sum_{i=1}^{n} v_{\text{ODE}}^2(t_i)} }
\end{equation}
where $\mathbf{v}_j(t_i), \mathbf{v}_{\text{ODE}}(t_i)$ denote the
$j^{\text{th}}$ observed trajectory in experimental data and the ODE
trajectory. The closer a trajectory is to the first-principle-based
ODE trajectories, the higher probability that the fluctuations result
from intrinsic noise (\textit{i.e.} the Brownian-type noise in
Eq.~\eqref{eq:nfkb_model-3}).

\section{Training the neural network to infer noise
  intensities in NF$\kappa$B dynamics}
\label{sec:infer_noise_nfkb}

We trained a neural network that took a group trajectories as the
input and then outputted inferred intrinsic noise intensity parameters
($\sigma_1$ and $\sigma_2$ in Eq.~\eqref{eq:nfkb_model-3}) for the
given group. Weights and biases in this neural network were optimized
by minimizing the mean squared error (MSE) loss:
\begin{equation}~\label{eq:pi}
    \text{MSE}(\Lambda)= \sum_{i=1}^m \|\sigma_{i}-\hat{{\sigma}}_i\|^2,
\end{equation}
where $\sigma_i$ are the ground truth noise intensity parameters
underlying the $i^{\text{th}}$ group of observed trajectories and
$\hat{\sigma}_i$ are the corresponding predicted parameters. Despite
the assumption of all cells sharing the same drift function (the same
underlying dynamics), different trajectories naturally arise from
intrinsic noise.
 
\begin{figure*}

  \centering
  \includegraphics[width=6in]{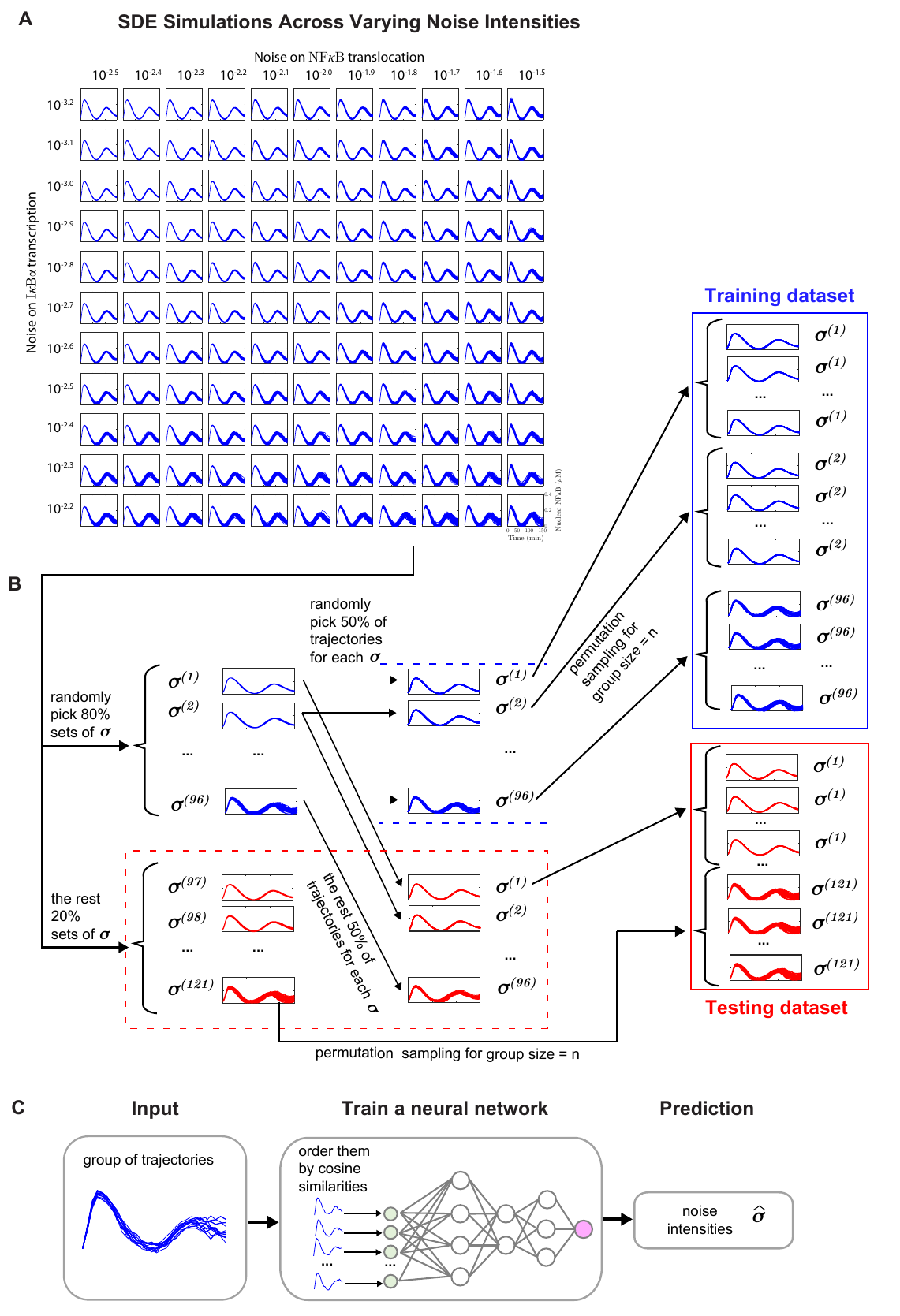}
  \caption{\textbf{Workflow of using neural network to infer noise
      intensities from a group of trajectories.} A. NF$\kappa$B SDE
    simulations under 121 different noise intensity settings.
    B. Schematic workflow for splitting the dataset into training and
    testing sets, where each group of trajectories is used to train
    and test the neural network for noise prediction.  C. Schematic of
    the training of a neural network to predict noise intensities for
    each group of trajectories.}
  \label{fig:noise_infer}
\end{figure*}

We adopted a neural network that takes a group of trajectories
($u_5(t)+u_{10}(t)$ in Eq.~\eqref{eq:nfkb_model-3}) under the same
noise intensity $(\sigma_1, \sigma_2)$ as the input and outputs the
inferred noise intensity $(\hat{\sigma}_1, \hat{\sigma}_2)$. The
training and testing sets are the same as those in Appendices
\ref{sec:simset_nfkb} and \ref{sec:train_nsde_nfkb}. The workflow for
splitting NF$\kappa$B SDE-simulated trajectories into training and
testing datasets is illustrated in Fig.~\ref{fig:noise_infer}A,
B. Specifically, out of the 121 combinations of noise intensities
(each containing 100 simulation trajectories), 20\% were designated
for the testing dataset pool. For the remaining 80\% noise
intensities, 50\% of the trajectories under each parameter were
randomly selected and added to the training dataset pool (indicated by
the blue box in Fig.~\ref{fig:noise_infer}B). The remaining 50\% of
trajectories from these parameter sets were used for the testing
dataset pool (red box in Fig.~\ref{fig:noise_infer}B). From both the
training and testing dataset pools, for each combination of noise
intensity, a group of trajectories were randomly sampled using a
permutation sampling approach to construct the training and testing
datasets (blue and red solid box in Fig.~\ref{fig:noise_infer}B).

The neural network used is equipped with an attention structure
\cite{vaswani2017attention} followed by a feed-forward structure of 2
hidden layers with 64 and 128 neurons in each layer, respectively,
where the attention mechanism is designed for assigning weights to
different trajectories in a group (sequenced by their similarities to
the deterministic ODE trajectory) (Fig.\ref{fig:noise_infer}C). The
hidden dimension of the attention structure in the query layer is 30
and the hidden dimensions of the attention structure in the key and
value layers are both 32.

To assess the impact of group size on the performance of the trained
neural network in predicting noise intensities, we tested group sizes
corresponding to 2, 4, 8, 16, and 32 trajectories per group. The
accuracy of the predictions was evaluated using the relative error
metric:
$$\frac{|\lg(\hat{\sigma}_1) - \lg(\sigma_1)|+|\lg(\hat{\sigma}_2) -
  \lg(\sigma_2)|}{|\lg(\sigma_1)| + |\lg(\sigma_2)|}.$$

Detailed steps on inferring noise intensities from a group of
trajectories are provided in Figs.~\ref{fig:noise_infer}A-C.

\end{document}